\documentclass[a4paper,11pt]{article}
\pdfoutput=1
\usepackage{jcappub}

\usepackage{graphicx}
\usepackage{color}
\usepackage[dvipsnames]{xcolor}
\usepackage{xspace}
\usepackage{multirow}
\usepackage{cleveref}
\usepackage[normalem]{ulem}
\usepackage{makecell}
\usepackage{appendix}

\newcommand{\tquote}[1]{``#1''}

\author[1,2]{Lorenzo Piga,}
\author[3]{Matteo Lucca,}
\author[4]{Nicola Bellomo,}
\author[5]{Valent\'i Bosch-Ramon,}
\author[6,7,8,9]{Sabino Matarrese,}
\author[6,7,8,10]{Alvise Raccanelli,}
\author[5,11]{and Licia Verde}

\affiliation[1]{Department of Mathematical, Physical and Computer Sciences, University of Parma, I-43124 Parma, Italy}
\affiliation[2]{INFN Gruppo Collegato di Parma, Parco Area delle Scienze 7/A, 43124, Parma, Italy}
\affiliation[3]{Service de Physique Th\'{e}orique, Universit\'{e} Libre de Bruxelles, C.P. 225, B-1050 Brussels, Belgium}
\affiliation[4]{Texas Center for Cosmology and Astroparticle Physics, Weinberg Institute, Department of Physics, The University of Texas at Austin, Austin, TX 78712, USA}
\affiliation[5]{Departament de Física Quàntica i Astrofísica, Institut de Ciències del Cosmos (ICCUB), Universitat de Barcelona (IEEC-UB), Martí i Franquès 1, 08028 Barcelona, Spain}
\affiliation[6]{Dipartimento di Fisica e Astronomia “Galileo Galilei”, Universit\`a degli Studi di Padova,
I-35131 Padova, Italy}
\affiliation[7]{INFN, Sezione di Padova, via F. Marzolo 8, I-35131 Padova, Italy}
\affiliation[8]{INAF - Osservatorio Astronomico di Padova, Vicolo dell’Osservatorio 5, I-35122 Padova, Italy}
\affiliation[9]{Gran Sasso Science Institute, Viale F. Crispi 7, I-67100 L’Aquila, Italy}
\affiliation[10]{Theoretical Physics Department, CERN, 1 Esplanade des Particules, CH-1211 Geneva 23, Switzerland}
\affiliation[11]{ICREA, Pg. Lluis Companys 23, E-08010 Barcelona, Spain}

\emailAdd{lorenzo.piga@unipr.it} 
\emailAdd{mlucca@ulb.ac.be} 
\emailAdd{nicola.bellomo@austin.utexas.edu}

\title{The effect of outflows on CMB bounds from Primordial Black Hole accretion}

\abstract{Should Primordial Black Holes (PBHs) exist in nature, they would inevitably accrete baryonic matter in their vicinity. In turn, the consequent emission of high-energy radiation could affect the thermal history of the universe to an extent that can be probed with a number of cosmological observables such as the Cosmic Microwave Background (CMB) anisotropies. However, our understanding of the accretion and radiation emission processes in the context of PBHs is still in its infancy, and very large theoretical uncertainties affect the resulting constraints on the PBH abundance. Building on state-of-the-art literature, in this work we take a step towards the development of a more realistic picture of PBH accretion by accounting for the contribution of outflows. Specifically, we derive CMB-driven constraints on the PBH abundance for various accretion geometries, ionization models and mass distributions in absence and in presence of mechanical feedback and non-thermal emissions due to the outflows. As a result, we show that the presence of such outflows introduces an additional layer of uncertainty that needs to be taken into account when quoting cosmological constraints on the PBH abundance, with important consequences in particular in the LIGO-Virgo-KAGRA observational window.}

\begin{document}

\hfill{\small ULB-TH/22-14, UTWI-11-2022}

\maketitle

\section{Introduction}

The $\Lambda$CDM model has become the standard cosmological model due to its remarkably accurate predictions of a variety of observables, ranging from the Cosmic Microwave Background (CMB) radiation~\cite{Aghanim2018PlanckVI} to the formation and evolution of large scale structures~\cite{Alam:2016hwk, Abbott:2017wau, Hildebrandt:2018yau}.\footnote{That is, of course, up to some tensions that recently emerged between the early-time inference and the late-time direct measurements of quantities such as the expansion rate of the universe today, $H_0$, and the amplitude of matter fluctuations, parameterized by~$\sigma_8$ (see, e.g., refs.~\cite{DiValentino2021Realm, Perivolaropoulos2021Challenges, Schoneberg2021Olympics, Abdalla:2022yfr} for recent reviews).} However, this phenomenological model does not provide any fundamental explanation for the nature of its \tquote{dark} components, i.e., cold Dark Matter (DM) and dark energy, despite the fact that they make up for the vast majority of the energy density of the universe today. Therefore, in order to find an origin for these fundamental, yet elusive, components of the universe, arguments invoking non-standard (beyond-$\Lambda$CDM) physics are often put forward. In particular, in the context of DM, Primordial Black Holes~(PBHs) have regained interest in recent years as potential DM candidates~\cite{bird:pbhasdarkmatter, clesse:pbhmerging, sasaki:pbhasdarkmatter} after the first LIGO detection of a Gravitational Wave (GW) signal emitted from a binary BH merger~\cite{LIGOScientific:2016aoc}.

PBHs are BHs that formed in the very early universe, much before the appearance of the first stars. Their existence was first suggested in the late `60s \cite{1967SvA....10..602Z, Hawking:1971ei, carr:pbhsformation, chapline:pbhformation} and they have since become one of the most popular candidates to make up for at least a fraction of the total DM content of the universe (see, e.g., refs.~\cite{Sasaki:2018dmp, Carr:2020gox, Carr:2020xqk, Villanueva-Domingo:2021spv} for recent reviews). Moreover it has been suggested that PBHs could also be the seeds for the formation of intermediate mass~\cite{2017ApJ...839L..13S} as well as super massive~\cite{Kohri:2014lza, Bernal:2017nec} BHs, which are believed to reside at the center of most galaxies. Other \textit{observational conundra} that are possibly solved by the potential presence of PBHs are discussed in refs.~\cite{Carr:2019kxo, Carr:2020xqk}. 

Despite these arguments in support of the presence of PBHs, many of their properties, such as their formation mechanism, abundance or mass distribution, are still unknown~\cite{Carr:2020xqk, Carr:2019kxo, Juan:2022mir}. As a consequence, existing constraints on the PBH abundance heavily rely on assumptions about the BH phenomenology. For instance, it is well known that PBHs, if existing, would inevitably accrete matter and convert a fraction of the energy released by the accretion into radiation. The resulting injection of high-energy photons into the cosmological photon bath would then affect the thermal history of the universe and, thus, observables that depend upon it, such as the CMB anisotropies. However, the impact of this energy injection on the CMB varies depending on the details of the accretion and energy injection mechanisms considered, for instance on the assumed geometry of the accretion~\cite{Ricotti:2007au, AliHaimoud2017Cosmic, Poulin2017CMB}, or on the properties of the environment surrounding the PBH~\cite{Mack:2006gz, Serpico:2020ehh}. Furthermore, the potential presence of magnetic fields and/or the excess of thermal energy in part of the accreted matter might lead to the formation of outflows, like winds or jets. Such outflows are expected to dilute the medium surrounding the BH reducing the total amount of accreted material, effect known as mechanical feedback (MF), and to be potential accelerators of non-thermal particles capable to effectively boost the total luminosity of the BH~\cite{Barkov:2012sj, Sadowski:2015jaa, Li:2019hfq, Bosch-Ramon:2020pcz, Bosch-Ramon:2022eiy}. Overall, the resulting CMB constraints on the PBH abundance cover approximately the~$\mathcal{O}(1-10^4)\ M_\odot$ mass range, with vastly different outcomes depending on the accretion and emission mechanisms \cite{AliHaimoud2017Cosmic, 2017JCAP...10..052B, Poulin2017CMB, Serpico:2020ehh}. These CMB anisotropy constraints are also complemented by bounds derived from CMB spectral distortions~\cite{AliHaimoud2017Cosmic}, galactic emission~\cite{gaggero:accretionconstraints, manshanden:accretionconstraint, hektor:accretionconstraintII}, interstellar gas heating~\cite{Takhistov:2021aqx}, and the 21-cm line~\cite{hektor:accretionconstraint, hutsi:accretionconstraint, mena:accretionconstraint}.

Yet, an accurate modelling of the accretion mechanism is fundamental to test the consistency between cosmological constraints and astrophysical hints in support of the PBH hypothesis. In fact, the range of PBH masses that would explain the origin of the LIGO-Virgo-KAGRA (LVK) observations lays between~$1$ and $100\ M_{\odot}$~\cite{LIGOScientific:2018mvr, Wong:2020yig, Franciolini:2021tla}, which is precisely the region of parameter space of relevance for CMB anisotropy constraints on matter accretion onto PBHs depending on the underlying assumptions of the accretion process~\cite{AliHaimoud2017Cosmic, Poulin2017CMB} (see, e.g., figure~2 of ref.~\cite{Carr:2019kxo} for a graphical representation). 

Currently, the two most popular models of accretion mechanism onto a PBH assume spherical~\cite{Horowitz:2016lib, AliHaimoud2017Cosmic} or disk~\cite{Poulin2017CMB} accretion, with the latter leading to a significantly higher radiation luminosity than the former, hence to more stringent bounds. Similarly, also the type of ionization mechanism at play in the vicinity of the PBH has been shown to be able to affect the radiation luminosity by orders of magnitude~\cite{AliHaimoud2017Cosmic}. 
However, in previous analysis the contribution from potentially present outflows has been neglected and it is therefore possible that their inclusion might alter the aforementioned conclusions. For instance, it is \textit{a priori} unclear whether the dominant effect of the outflows is the reduction of the accretion rate via MF or the enhancement of the radiation luminosity due to the non-thermal emission of photons. The relative importance of these effects is very likely to depend not only on the PBH mass, but also on other assumptions regarding the geometry of the accretion, the ionization model and the characteristics of the outflows.

In this work we investigate these open questions by constructing an accretion model that accounts for both MF and non-thermal outflows, building on the analysis previously carried out in refs.~\cite{Bosch-Ramon:2020pcz, Bosch-Ramon:2022eiy}. We then study their impact on the thermal history of the universe and the CMB anisotropies in the context of both accretion geometries and ionization models, as well as for various PBH mass distributions. As a result, we find that the large theoretical uncertainties underlying the modelling of the outflows significantly enlarge those already present due to the other variables, leading to upper bounds on the PBH mass that can vary from approximately 0.01 to 100 $M_\odot$ (assuming PBHs make all of the DM) depending on the chosen accretion model and outflow properties. This in turn bears important consequences for the LVK observability window. The work presented here shows that a more robust and complete modelling of PBH accretion is necessary to draw meaningful conclusions on PBHs being realistic DM candidates and progenitors of the LVK detected events.

The paper is organized as follows. In section~\ref{sec: state_art} we briefly introduce the generalities of PBH accretion and outline the current status of the field, while in section~\ref{sec:mechanical_feedback} we present the outflow model considered here in its details, adopting and extending the main results of~\cite{Bosch-Ramon:2020pcz, Bosch-Ramon:2022eiy}. In section~\ref{sec: ther_hist} we succinctly explain how the accretion process affects the thermal history of the universe. In section~\ref{subsec: num} we discuss the numerical implementation of the accretion models, while in section~\ref{sec:cmb_analysis_constraints} we calculate the resulting CMB anisotropy constraints for monochromatic and extended PBH mass functions, and discuss their implications for the LVK observed mass range. Finally, we conclude in section~\ref{sec: concl} with a summary of our analysis and closing remarks.


\section{State of the art of PBH accretion}
\label{sec: state_art}

A PBH is moving supersonically in a homogeneously distributed gas accretes matter though an accretion column forming opposite to the direction of motion. In this scenario the accretion at the BH horizon is conveniently described by the Bondi-Littleton rate~\cite{hoyle:accretionrate, bondi:accretionrate, Bondi:1952ni}
\begin{align}\label{eq: m_dot}
    \dot{M}_\mathrm{PBH} = 4\pi \rho_\infty\, v_{\rm rel}\, \lambda\, r_B^2\,,
\end{align}
where~$\rho_\infty$ is the mass density of the gas far away from the point mass, $v_{\rm rel}$ is the relative velocity between the PBH and the gas, $\lambda$ is the dimensionless accretion rate (which takes into account deviations from the idealised Bondi scenario due to the presence of e.g., pressure, viscosity, radiation feedback, MF, etc.), and~$r_B$ is the Bondi radius, which characterizes the size of the BH sphere of influence and is defined as ${r_B=GM_\mathrm{PBH}/v_{\rm rel}^2}$. The consequent radiation luminosity of the system is typically parametrized as
\begin{align}\label{eq: luminosity}
    L_\mathrm{rad} = \epsilon \, \dot{M}_\mathrm{PBH}\,,
\end{align}
where~$\epsilon$ is a dimensionless parameter that accounts for the radiation efficiency of the accretion process.\footnote{In equations~\eqref{eq: m_dot}-\eqref{eq: luminosity} we have assumed $c=1$, convention that we will apply throughout the manuscript.}

While the PBH mass is a free parameter of the model, quantities like $v_{\rm rel}$, $\lambda$ and $\epsilon$ are determined by the properties of the environment close the PBH, such as, for instance, the temperature profile of the infalling gas, as well as by the geometry of the accretion or the details of the energy emission. Although there is no consensus on the exact form that these quantities should take, several ideas have been proposed so far in the literature (see e.g., refs.~\cite{AliHaimoud2017Cosmic, Poulin2017CMB}). In order to set the stage for the accretion model described in section~\ref{sec:mechanical_feedback}, where we focus more closely on the role of outflows, in this section we briefly review the state of the art of PBH accretion. Specifically, we focus on the role of the relative velocity between PBH and surrounding gas in section~\ref{subsec:vel}, on that of the gas ionization model in section~\ref{subsec:ion} and on that of the geometry of the accretion in section~\ref{subsec:geom_acc}, since these are the main aspects that affect the final accretion radiation luminosity in absence of outflows. The possible impact of other physical effects is discussed in section~\ref{subsec:add_eff}.


\subsection{Accounting for PBH velocities}
\label{subsec:vel}

Constraints on the PBH abundance have been reported for different choices of the PBH velocities. Since the different assumptions can change the resulting constraints by orders of magnitude, we analyze and motivate here the approach employed in this work.

The first possibility is to consider that PBHs are at rest or moving at sub-sonic speed, hence the only relevant velocity is that of the surrounding medium, i.e., the speed of sound far away from the accretion region ${c_{s,\infty}^2=\gamma v^2_B = \gamma P_\infty/\rho_\infty}\,$, where~$\gamma=5/3$ is the polytropic equation of state index for an ideal monoatomic gas and~$v_B=\sqrt{P_\infty/\rho_\infty}$ is the Bondi velocity. Therefore, in the \tquote{$v_{\rm PBH}=0$} scenario, where $v_{\rm PBH}$ is the PBH proper velocity, we have
\begin{align}\label{eq: v_reff_1}
	v_{\rm rel}=c_{s,\infty}\,.
\end{align}
However, under the assumption that the PBHs are the DM, PBHs are expected to be super-sonic given that their velocity would effectively be the DM-baryon linear relative velocity~$v_L$, at least on large scales.\footnote{Following ref.~\cite{AliHaimoud2017Cosmic}, we only consider the large-scale linear velocity of the PBHs, neglecting the effects of the small-scale non-linear contribution from PBH clustering. As shown in ref.~\cite{Inman:2019wvr}, this approach is accurate for $z\gtrsim 100$, which is also the redshift range CMB data are the most sensitive to (see, e.g., ref.~\cite{Slatyer2016General}). We therefore do not expect this approximation to affect our results significantly.} In this case the relative velocity reads
\begin{align}\label{eq: v_reff_2}
	v_{\rm rel}=\sqrt{c_{s,\infty}^2+v_{\rm PBH}^2}\,.
\end{align}

Since velocities are stochastic variables, the radiation luminosity of a PBH population is in fact a velocity-averaged\footnote{For the cases presented in this work velocities are assumed to be Gaussian distributed.} luminosity~$\langle L_{\rm rad} \rangle$. So far two different averaging procedures have been presented in the literature \cite{AliHaimoud2017Cosmic}. In the first one, which we refer to as \tquote{approximated} average in the rest of the paper, the relative velocity is approximated as
\begin{align}\label{eq: v_reff_3}
	v_{\rm rel}\simeq\sqrt{c_{s,\infty}\langle v_{\rm PBH}^2 \rangle^{1/2}},
\end{align}
with 
\begin{equation}
    \langle v_{\rm PBH}^2 \rangle^{1/2} = \langle v_L^2 \rangle^{1/2} = \text{min}[1,(1+z)/10^3]\times 30\ \mathrm{km/s},
\end{equation}
$z$ being the redshift, and this result is used in equation~\eqref{eq: m_dot} to obtain the radiation luminosity of the population. This approximation is well motivated in the cases where the radiation luminosity is proportional to~$\dot{M}_{\rm PBH}^2 \propto (c_{s,\infty}^2+v_{\rm PBH}^2)^{-3}$, however in different scenarios its validity has to be assessed on a case-by-case basis. The approximation in equation~\eqref{eq: v_reff_3} holds for all redshifts below $z\simeq10^4$ and can therefore be safely applied for all times of interest for this work.

The second, more exact and general way of averaging over relative velocities is to consider that a transformation of the form $c_{s,\infty}^2\to c_{s,\infty}^2+v_{\rm PBH}^2$ (which is the same as going from equation~\eqref{eq: v_reff_1} to equation~\eqref{eq: v_reff_2}) corresponds to a transformation of the gas temperature far away from the PBH of the form
\begin{align}\label{eq: v_reff_4}
	T_{\infty} \to T_{\infty} + \frac{m_p\, v_{\rm PBH}^2}{\gamma(1+\bar{x}_e)}\,,
\end{align}
where $m_p$ is the proton mass and $\bar{x}_e$ is the background free electron fraction.\footnote{This transformation follows from the fact that $c_{s,\infty}^2=\gamma T_\infty (1+\bar{x}_e) /m_p$, and hence $T_\infty = m_p c_{s,\infty}^2/(1+\bar{x}_e)\gamma $.} In this case we express~$v_{\rm rel}$ in terms of $T_{\infty}$ and average the resulting radiation luminosity over the distribution of the relative velocities to obtain the total radiation luminosity $\langle L_{\rm rad} \rangle$. In the following we refer to this approach as \tquote{exact} average.

\begin{figure}[t]
	\centering
	\includegraphics[width=0.48\columnwidth]{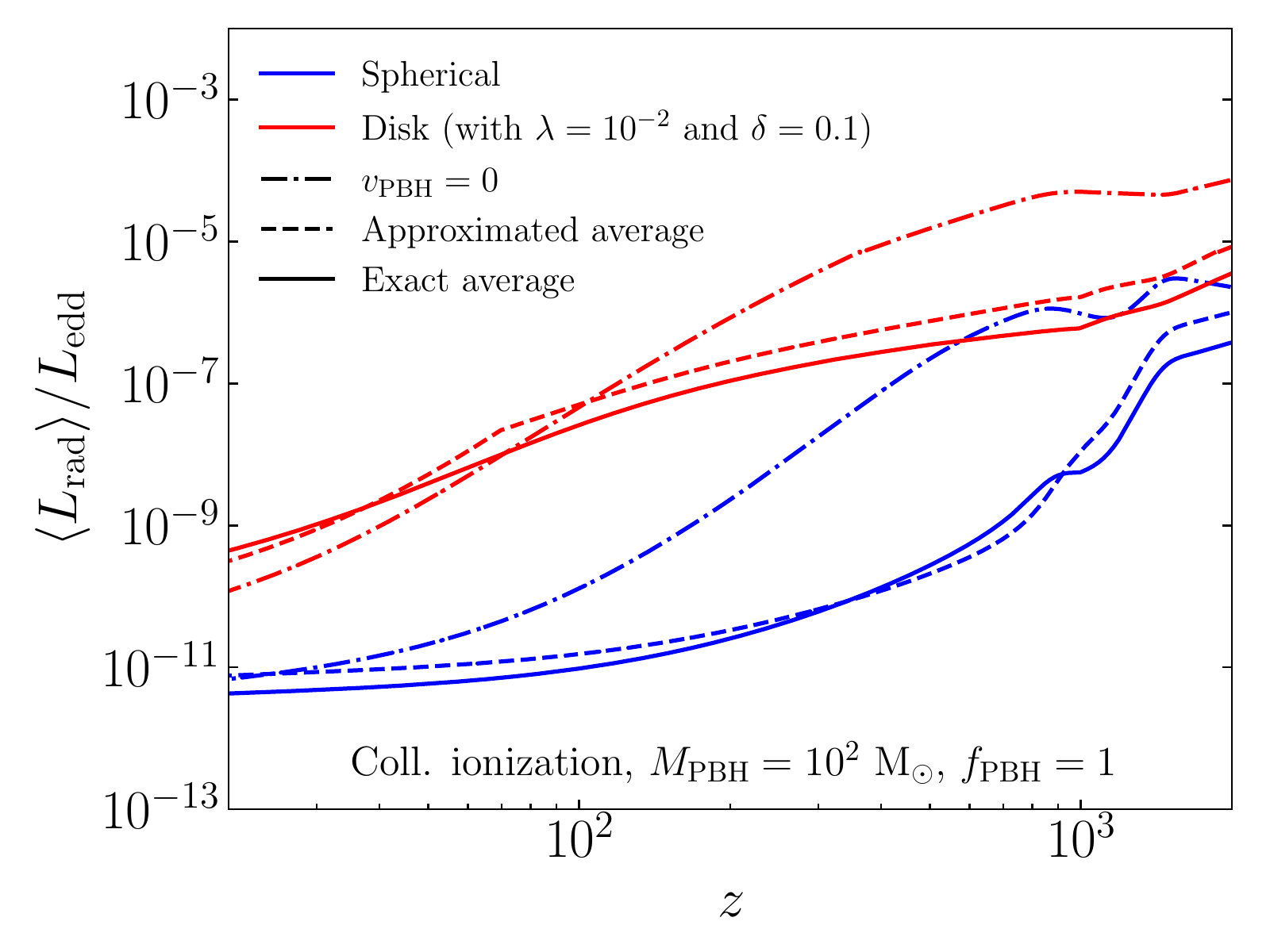}	\includegraphics[width=0.48\columnwidth]{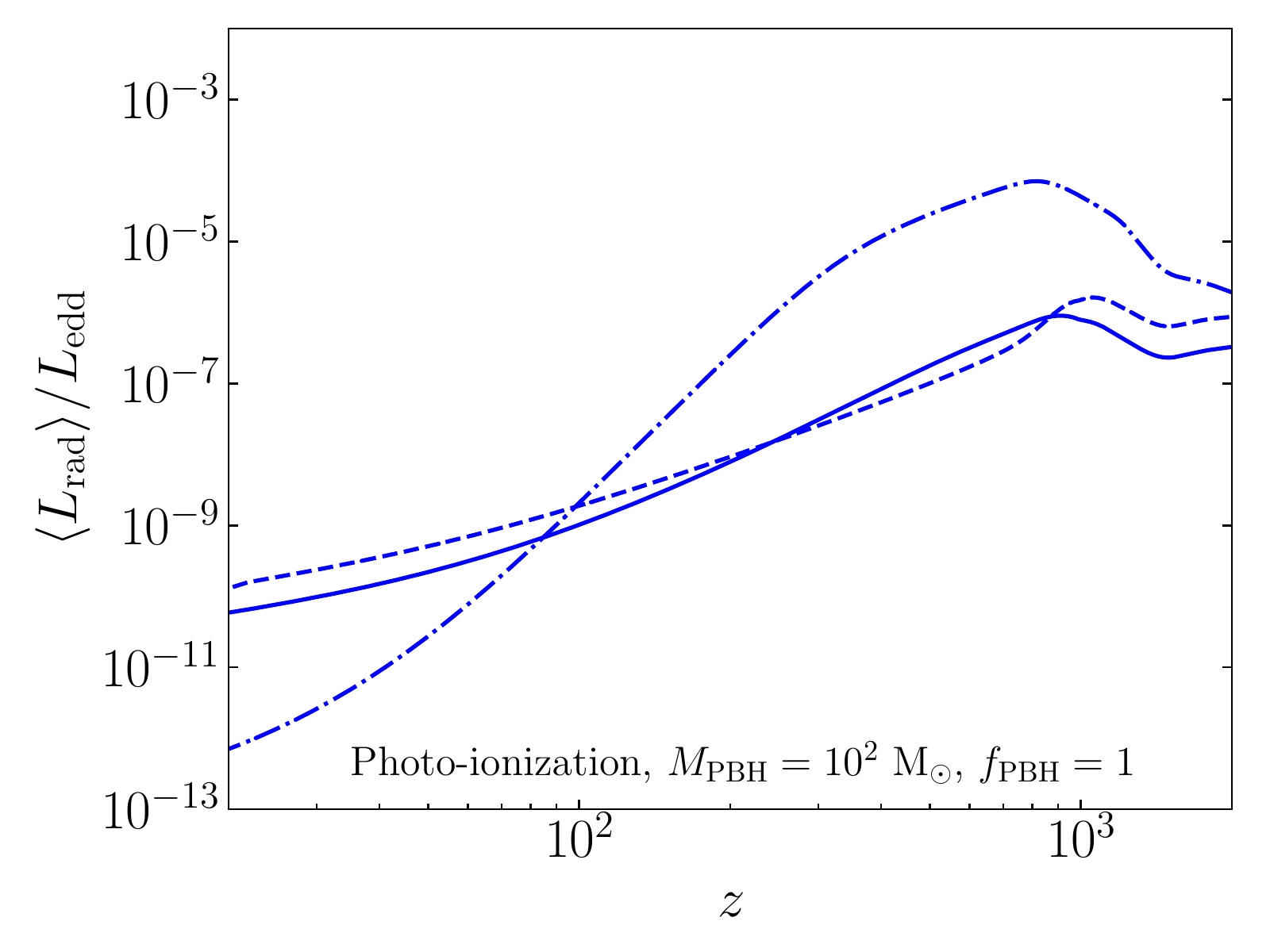}
	\caption{Graphical illustration of the impact that the three possible choices for the computation of the relative velocity between PBH and surrounding matter have on the final luminosity of the system, for both the collisional (left) and photo-ionization (right) models. As term of reference, we compare the average radiation luminosity $\langle L_{\rm rad} \rangle$ to the Eddington luminosity~$L_\mathrm{edd}$.}
	\label{fig: v_rel}
\end{figure} 

We show in figure~\ref{fig: v_rel} the effect that these three choices for the relative velocity ($v_{\rm PBH}=0$, approximated and exact average) have on the averaged total luminosity~$\left\langle L_\mathrm{rad} \right\rangle$. In the figure we consider the two types of ionization mechanisms described in section~\ref{subsec:ion} as well as both the spherical and disk accretion scenarios detailed in section~\ref{subsec:geom_acc}, for the representative case of $M_{\rm PBH}=10^2\, M_\odot$ and $f_{\rm PBH}=1$, where $f_{\rm PBH}=\rho_{\rm PBH}/\rho_{\rm DM}$ is the fractional PBH energy density normalized to the DM abundance.\footnote{Although not explicitly present in the definition of the radiation luminosity, $f_{\rm PBH}$ affects its value since $L_{\rm rad}$ depends on the background temperature of the accreted matter, which is in turn sensitive to the total amount of injected energy and hence to $f_{\rm PBH}$, as discussed in section~\ref{sec: ther_hist}. We will therefore explicitly report the employed value of $f_{\rm PBH}$ in all figures presented in the manuscript.} As can be clearly seen from the figure, the two methods to account for $v_{\rm rel}$ when $v_\mathrm{PBH}\neq 0$ are overall almost equivalent, although differences (roughly of order two for the cases considered in the figure) can appear in particular for the disk accretion scenario, as in this case the radiation luminosity is not necessarily proportional to $\dot{M}^2_\mathrm{PBH}$. Moreover, we find that the exact approach is the most conservative. The figure further highlights the importance of including the PBH proper velocity, since the predictions of equation~\eqref{eq: v_reff_1} always significantly deviate (even by orders of magnitude) from the exact approach. Therefore, for the sake of generality and being conservative, henceforth (unless stated otherwise) we always employ the exact averaging procedure to account for relative velocities, as done for instance in ref.~\cite{AliHaimoud2017Cosmic}. 


\subsection{Ionization models}
\label{subsec:ion}

In the scenarios of interest for this work, the radiation luminosity of an accreting PBH mostly comes from free-free radiation originated in the region close to the Schwarzschild radius~\cite{AliHaimoud2017Cosmic}. The energy of the photons emitted in that region depends on the plasma temperature close to and around the compact object, which in turn depends on the dynamics of the infalling gas. Therefore, an accurate estimate of the temperature profile of the accreting region is necessary to determine the PBH radiation luminosity.

Before recombination, when the medium is already ionized, the temperature increases adiabatically the closer the gas gets to the BH. After recombination, when the medium far from the compact object becomes neutral, two scenario are possible, depending on the relative importance of two potentially competing effects: medium compression, which increases the temperature adiabatically, and ionization, which reduces the temperature proportionally to the energy lost in the ionization process and hence to the amount of initial neutral gas. In the first scenario the temperature of the infalling gas increases until it reaches~$T_\mathrm{ion}\sim 10^4\ K$, then it stops increasing until all the gas gets ionized (since the balance between compression and ionization is maintained), and finally it resumes increasing. In this scenario the neutral gas is ionized through collisions with the free electrons and it is therefore referred to as ``collisional ionisation''. On the other hand, in the second scenario the radiation produced near the Schwarzschild radius is intense enough to directly photo-ionize the infalling gas before it reaches the ionization temperature~$T_\mathrm{ion}$. Therefore in this scenario, which is conveniently called ``photo-ionization'', the temperature increases monotonically. In both scenarios radiative cooling is not expected to be relevant on scales close to the PBH.

Given that determining a realistic profile of the temperature around the PBH is very challenging and would require dedicated numerical simulations, we follow the simplifying approach of ref.~\cite{AliHaimoud2017Cosmic}: we consider these two limiting scenarios of purely collisional or photo- ionization and report our results for both of them, knowing that they would bracket the \tquote{realistic} results.


\subsection{Geometry of the accretion mechanism}
\label{subsec:geom_acc}

The geometry of the accretion plays a crucial role in determining both the dimensionless accretion rate~$\lambda$ and the radiation efficiency~$\epsilon$. As in the previous section, also in this case we treat the spherical and disk accretion scenarios as the limits that encompass the \tquote{true} results.

\begin{figure}[t]
    \centering
    \includegraphics[width=0.48\textwidth]{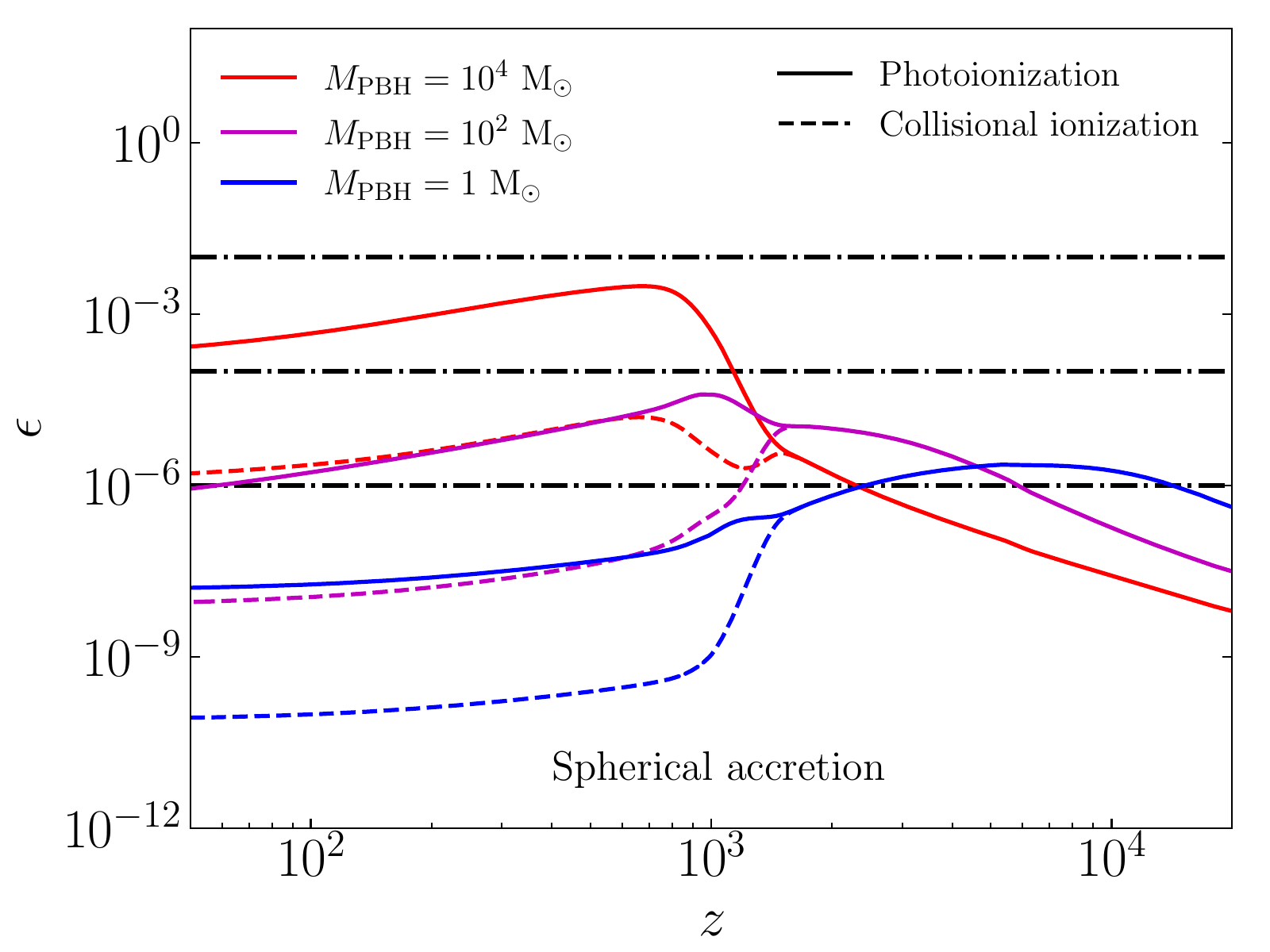}
    \includegraphics[width=0.48\textwidth]{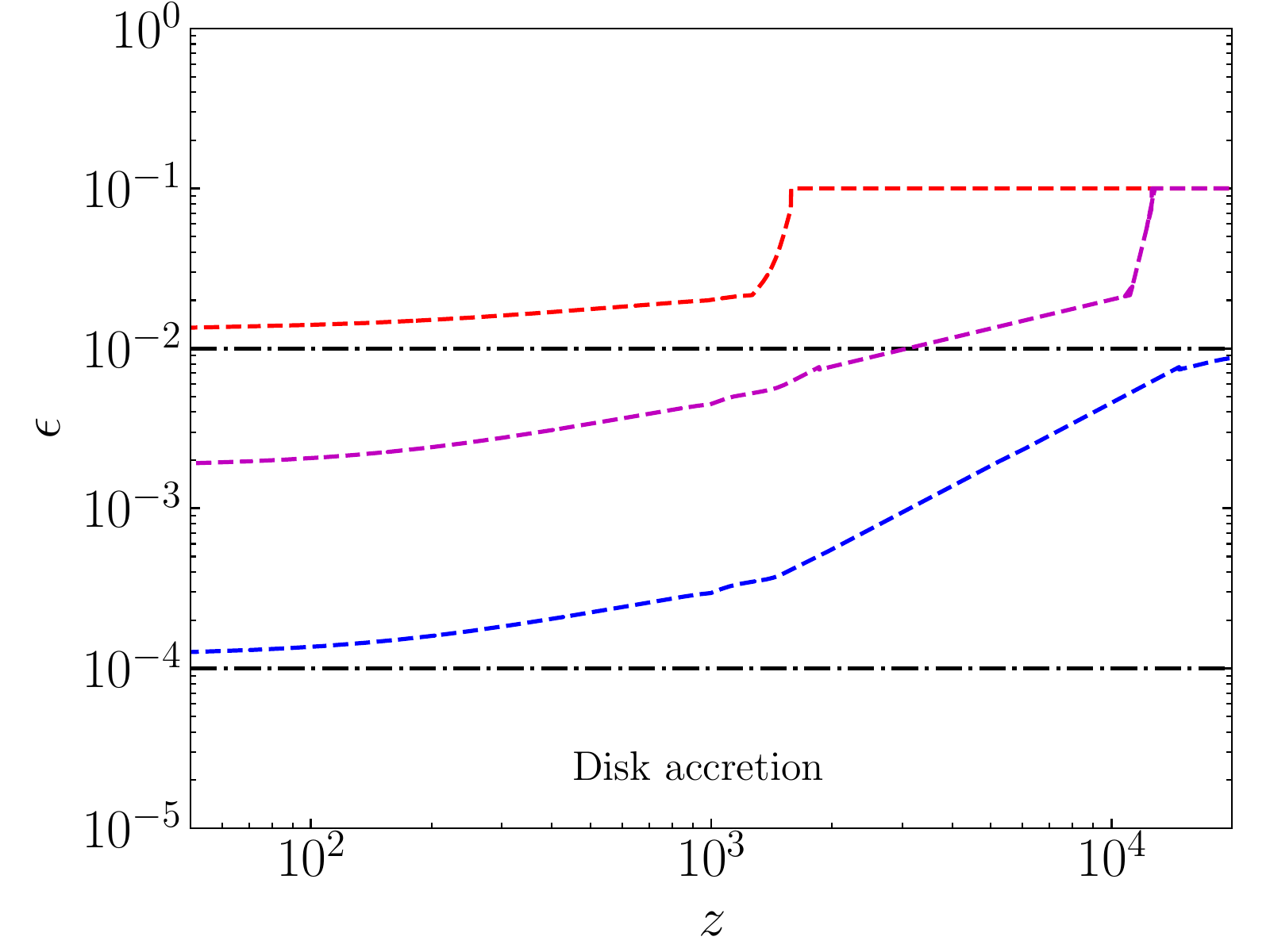}
    \caption{Graphical illustration of the dependence of the emission efficiency $\epsilon$ on the geometry of the accretion (spherical on the left, disk on the right) as well as on the ionization mechanism (photoionization as solid lines, collisional ionization as dashed lines) for different choices of the PBH mass. In all cases we assume $f_{\rm PBH}=1$. The horizontal dashed-dotted lines represent the benchmark outflow efficiencies $\epsilon_{\rm non-th}$ discussed in section \ref{subsec: imp_lum}.}
    \label{fig: plot_eps}
\end{figure}

In the spherical accretion scenario, most of the calculations can be performed semi-analytically (see, e.g., ref.~\cite{AliHaimoud2017Cosmic} for a recent treatment of the problem). The accretion efficiency can be expressed as a function of the Compton drag and cooling rate by CMB photons, finding limiting cases where these rates are either very efficient (isothermal accretion with $\lambda\simeq1.1$) or negligible (adiabatic accretion with $\lambda\simeq0.1$). It is also possible to study the evolution of the dimensionless accretion rate depending on the mass of the PBH and on the epoch of the universe, as can be seen in e.g., figure~4 of ref.~\cite{AliHaimoud2017Cosmic}. Similarly, the radiation efficiency reduces to a function of the mass accretion rate and the temperature of the gas close to the PBH, which in turn depends on the ionization mechanism as discussed in the previous section. Typical values for the radiation efficiency in the spherical case are displayed in the left panel of figure~\ref{fig: plot_eps} for difference choices of the PBH mass and both ionization models.

The other accretion mechanism we consider assumes the formation of a disk around the PBH~\cite{Poulin2017CMB}. Given that PBHs are expected to form binary systems already at early times~\cite{Sasaki:2016jop}, it seems natural to expect that the gas around the PBH experiences tidal forces that lead to the formation of a disk. Unfortunately, the details of disk accretion are very difficult to determine semi-analytically and only educated estimates have been used so far in the literature to determine quantities such as $\lambda$ or $\epsilon$. Following ref.~\cite{Poulin2017CMB}, we fix as (arbitrary) benchmark value for the dimensionless accretion rate~$\lambda=0.01$, which has been argued to roughly take into account the role of viscosity and outflows close to the BH.\footnote{The role of outflows in the works upon which ref.~\cite{Poulin2017CMB} is based is different than what is considered in section~\ref{sec:mechanical_feedback} (i.e., MF). In that context, outflows play a role only close to the BH, on scales of order approximately~$10-100$ times the Schwarzschild radius, by removing material close to the compact object, thus reducing the accretion rate onto the BH. Hence they act at small scales, not at large scales reducing the available gas beyond the PBH sphere of influence. Moreover, in those works, the luminosity efficiency is boosted because of the increased energy transferred to electrons in the vicinity of the BH, not because of the acceleration of particles.} Similarly, we employ a parametric form of the radiation efficiency $\epsilon$ provided by ref.~\cite{Xie:2012rs} (see in particular equation~(11) and table 1 therein). However, there is an order-of-magnitude uncertainty in the estimate of the energy in electrons, characterized as the fraction of energy present in ions, which we assume here to have the representative value of $\delta=0.1$, as done in ref.~\cite{Poulin2017CMB}. The redshift evolution of~$\epsilon$ in the disk accretion scenario for these benchmark choices is displayed in the right panel of figure~\ref{fig: plot_eps}. Note that these values are up to six orders of magnitude higher than those typical of the spherical accretion scenario. The figure also highlights that, while in the spherical accretion scenario~$\epsilon\propto \dot{M}_{\rm PBH}/L_{\rm edd}\propto M_{\rm PBH}$, in the disk accretion scenario~$\epsilon\propto (\dot{M}_{\rm PBH}/L_{\rm edd})^a\propto M_{\rm PBH}^a$ where~$a\sim 0.3-0.6$ is a parameter fitted to match numerical simulations of ref.~\cite{Xie:2012rs}, and hence the radiation efficiency scales differently as a function of the PBH mass in the two cases.

In summary, the unknown geometry of the accretion introduces an enormous source of uncertainty - potentially of orders of magnitude - in assessing the final radiation luminosity. This uncertainty is to be added on top of the uncertainty due to the ionization processes, although so far the disk accretion setup has been analyzed only in the context of collisional ionization. Due to these not completely understood physical processes, the constraints on the abundance of PBHs can vary substantially depending on the underlying choices made to describe the accretion process. Additional effects (discussed in detail in section \ref{subsec:add_eff}) only further add to this uncertainty budget.


\subsection{Additional effects}
\label{subsec:add_eff}

Besides these main factors, the accretion process can potentially be affected by several other contributions that have been considered in the literature. In this section we briefly outline a number of them, highlighting the regimes where they become relevant or can be neglected.

\begin{itemize}
    \item \textit{The role of the Hubble expansion}. This effect acts on the accretion process as a form of viscosity and introduces an additional redshift dependence of the background energy density and sound speed~\cite{Ricotti:2007jk}. However, these contributions become important only for~$M_{\rm PBH}\gtrsim 10^4\ M_\odot$, threshold above which the quasi-steady flow approximation breaks down and so does the underlying mathematical setup. Therefore, in particular since in our analysis we focus on the LVK mass range, we do not consider masses above~$10^4\ M_\odot$.
    \item \textit{The gravitational effect of the gas surrounding the PBH}. The effect of self-gravity of the accreting gas can be neglected as long as the PBH mass is much larger than that of the surrounding medium, which is always the case for $M_{\rm PBH}\lesssim 10^5\ M_\odot$ \cite{Ricotti:2007jk} (as for these masses the PBH can only attract a relatively low amount of matter) and therefore also for the range of PBH masses considered in this work. The same kind of consideration applies also for the gravitational impact of the background matter on the PBH (which could in principle affect the motion of the PBH, effect known as dynamical friction) as long as one considers PBH masses below~$10^4\ M_\odot$~\cite{Serpico:2020ehh, Takhistov:2021aqx}.
    \item \textit{The role of super-Eddington accretion}. Another possibility to consider for very massive PBHs is for them to accrete at a super-Eddington rate (see, e.g., refs.~\cite{Mayer:2018vrr, Brightman:2019nwc, Takeo:2019uef} and references therein for recent reviews). The Eddington accretion rate is the limit at which the gravitational force of the BH and the radiation pressure compensate (under the assumption of spherical symmetry), defining therefore the threshold for the maximum luminosity of the object. However, in practice, this limit can be exceeded for instance within proto- and massive galaxies at redshifts~$z\lesssim 20$, leading to periods of enhanced but variable luminosity. For spherical accretion, the transition to super-Eddington accretion is believed to happen for PBH masses roughly above~$10^4\ M_\odot$, although in more realistic situations this value would strongly depend on the characteristics of the surrounding environment and on the geometry of the accretion (see, e.g., ref.~\cite{Takeo:2017qbj}). Nevertheless, since this is a mainly low-redshift effect, our CMB bounds are largely unaffected by this source of uncertainty.
    \item \textit{The role of large-scale structure formation and DM halos}. Once large-scale structures start to form at redshifts~$\mathcal{O}(10)$, the non-trivial dynamics of PBHs and their interaction with the medium requires numerical simulations to be properly assessed~\cite{inman:pbhlateuniverse, boldrini:pbhlateuniverse, liu:pbhlateuniverseI, liu:pbhlateuniverseII}, even if semi-analytical arguments can be made regarding how the accretion proceeds (see, e.g., ref.~\cite{deluca:pbhevolution, DeLuca:2020fpg}). Furthermore, in the scenario where PBHs represent just a sub-dominant component of the DM and the rest is described by some new beyond-the-Standard-Model particle, DM halos are expected to form around the PBHs \cite{Adamek2019WIMPs}. In this case, the formation of DM halos and of large-scale structures more in general can boost both the accretion of material~\cite{Mack:2006gz, Ricotti:2007jk, Ricotti:2007au} and the PBH proper velocities \cite{Poulin2017CMB}. While the former effect enhances the radiation luminosity~\cite{Serpico:2020ehh}, the latter reduces it~\cite{DeLuca:2020fpg}. However, since these competing effects are strongly reliant on the assumptions made to describe the relevant late-time physics, we will neglect them in the current analysis for sake of clarity and leave their inclusion for future work.
    \item \textit{PBH mutual interactions.} In the absence of large primordial non-Gaussianity, PBHs do not cluster over cosmic times~\cite{alihaimoud:pbhclustering} (see e.g., ref.~\cite{deluca:initialclustering} for further discussions) and PBH evolution can be effectively considered in isolation for masses below~$10^4\ M_\odot$~\cite{AliHaimoud2017Cosmic}. Furthermore, although early works suggested the existence of a large number of PBH binaries forming at early times~\cite{alihaimoud:pbhmergerrate}, these are in contradiction with more recent numerical simulations~\cite{jedamzik:earlybinariesmergerrateI, jedamzik:earlybinariesmergerrateII}. Hence, PBH mutual interactions can be safely neglected in the mass range considered in this work.
    \item \textit{The role of the PBH spin}. The radiation emitted by fastly rotating PBHs is enhanced by a factor of order ten percent with respect to the non-rotating case~\cite{1974ApJ...189..343S}. However, the majority of Kerr PBHs are realistically expected to have a dimensionless Kerr parameter much lower than unity~\cite{Chiba:2017rvs, deluca:pbhspindistribution}, especially in the scenarios where PBHs form from the collapse of large primordial density fluctuations (see, e.g., refs.~\cite{Harada:2017fjm, Kuhnel:2019zbc, deFreitasPacheco:2020wdg} for possible exceptions). Hence this effect can be neglected for the purpose of this work. 
    \item \textit{The effect of inhomogeneuos energy injections}. Recently, it has also been shown that the impact of PBH accretion on the photon bath can be very sensitive to the inhomogeneous spatial distribution of PBHs~\cite{Jensen:2021mik}. Nevertheless, since further important theoretical and numerical steps will be necessary to fully understand this effect, as also pointed out in the reference, we leave its inclusion in the analysis for future work.
\end{itemize}


\section{The role of outflows}
\label{sec:mechanical_feedback}

Another key physical effect that has been neglected in many of the previous analyses is the presence of outflows and their effect on large scales, where hereafter in this context ``large'' implies approximately the PBH scale of influence~$r_\mathrm{acc} \simeq 2GM_\mathrm{PBH}/v^2_\mathrm{PBH}=2 r_B$. Recent analytical estimates and numerical simulations~\cite{Bosch-Ramon:2020pcz, Bosch-Ramon:2022eiy} suggest that even relatively weak jets and winds, i.e., outflows with different degree of collimation, might be able to sweep away some of the material around the PBH, thus reducing the flow of gas entering the BPH sphere of influence. Therefore, if present and carrying as little power as $10^{-6}\dot{M}_{\rm PBH}$~\cite{Bosch-Ramon:2020pcz}, outflows could reduce the available material for accretion through this mechanical feedback effect. Despite the fact that the magnetic field in the primordial gas and the spin of the PBH are likely to be small, even pure hydrodynamical effects could drive outflows in a non-rotating black hole under small anisotropies of the accretion structure \citep{Bosch-Ramon:2020pcz,Aguayo-Ortiz2019,Tejeda2020,Waters2020}. Taking this and the uncertainties on accretion physics at very high $z$ into account, the role of mechanical feedback deserves to be considered. 

Furthermore, if outflows are produced during the accretion process, they can also potentially accelerate particles that can radiate and inject an additional amount of energy into the thermal bath. Such radiation from the outflows can compete with that produced in the accretion process itself, however, only for outflows that are relatively powerful. Therefore, outflows with a rather low power in terms of $\dot{M}_{\rm PBH}$ could still produce mechanical feedback but generate negligible amounts of radiation. These two phenomena, analyzed separately in sections~\ref{subsec: imp_acc} and~\ref{subsec: imp_lum}, are in fact competing effects: while decreasing the accretion rate intrinsically reduces the PBH radiation luminosity, such loss of luminosity is potentially compensated by the energy deposited into the medium by non-thermal particles accelerated by the outflows. 

In this work we analyze the impact outflows potentially have on the two different accretion geometries presented in section~\ref{subsec:geom_acc}. As explained before, these models have to be understood as limiting cases, presented to illustrate where the real physical scenario potentially lays in the PBH parameter space. Therefore, even if strictly speaking in the perfectly spherically symmetric scenario we cannot have outflows, it is still useful to consider that case as the benchmark reference for the scenario where some very low amount of angular momentum is present, which is expected in a cosmological scenario with small inhomogeneities, as outlined in ref.~\cite{Bosch-Ramon:2020pcz}.


\subsection{Impact on the accretion rate}
\label{subsec: imp_acc}

As discussed in ref.~\cite{Bosch-Ramon:2020pcz}, many different mechanisms promote the formation of outflows as, for instance, the existence of a magnetic field and/or the presence of some accreted gas with an excess of thermal energy. A (small) magnetic field can be generated by the Biermann battery mechanism, in which case the magnetorotational instability could enhance the magnetization and the turbulence, creating the required conditions for outflow formation~\cite{2018MNRAS.479..315S}. In the presence of magnetic fields, outflows and jets can be fed by the rotational energy of the PBH or by the inner regions of an accretion disk~\cite{1977MNRAS.179..433B,1982MNRAS.199..883B}, even when the accretion is quasi-spherical~\cite{2012MNRAS.421.1351B, 10.1111/j.1365-2966.2012.22029.x,2021MNRAS.504.6076R}. On the other hand, the production of an outflow from an excess of thermal energy in the gas is related to the formation of a thick and radiatively inefficient accreting disk, which forms winds through thermal pressure gradients. These outflows are expected to be broad and largely independent of the details of the magnetic field~\cite{Sadowski:2015jaa}.

As demonstrated in ref.~\cite{Bosch-Ramon:2020pcz}, once launched, outflows can overcome the ram pressure of the infalling gas and escape the PBH sphere of influence even when they are relatively weak and non-relativistic. Once outflows reach large scales, they compress and heat the medium by depositing energy and momentum into the gas. This phenomenon, called mechanical feedback (MF), is potentially responsible for a decrease in the accretion rate because the material is effectively heated and swept away from the PBH, implying that the amount of material that is available at scales of order~$r_\mathrm{acc}$ is reduced with respect to the standard case presented in section~\ref{sec: state_art}.

In order to account for this new large-scale effect, we modify the PBH accretion rate in equation~\eqref{eq: m_dot} by introducing a new fractional rescaling $f_\mathrm{LS}$, so to obtain
\begin{equation}
    \dot{M}_\mathrm{PBH} = 4\pi\rho_\infty v_{\rm rel}\, \lambda\, f_\mathrm{LS}\, r_B^2\,.
\label{eq:mdot_mechanical_feedback}
\end{equation}
This new parameter specifically accounts for physical effects happening at large scales. Therefore, in the case with no MF, we have by definition that~$f_\mathrm{LS}\equiv 1$ and we recover the cases presented in section~\ref{sec: state_art}. Conversely, when MF is present, $f_\mathrm{LS}$ is expected to be lower than unity with a value that depends on the outflow ejected material velocity~$v_\mathrm{o}$, the angle between the outflow and the PBH velocity~$\theta$ and outflow half-opening angle~$\chi$ (even if this latter dependence is weaker than the first two) \cite{Bosch-Ramon:2022eiy}. 

Numerical simulations in 3D~\cite{Bosch-Ramon:2022eiy} suggest that a conservative choice for the fractional rescaling benchmark value (after averaging over all the possible values of the outflow orientation $\theta$) is~$f_\mathrm{LS}=0.1$. Despite implying an order-of-magnitude reduction of the accretion rate, this choice is still cautious since phenomena such as the potential fast variation of the outflow orientation may further reduce the value of~$f_\mathrm{LS}$ and hence of the accretion rate (as discussed in \cite{Bosch-Ramon:2022eiy}). Moreover, once PBHs become trans- or sub-sonic, the value of the fractional rescaling may be further reduced since the system becomes more axi-symmetric, and the accumulated internal energy and thus pressure within the outflow-medium interaction region can diminish accretion further (see, e.g., \citep{Zeilig-Hess2019,Bosch-Ramon:2020pcz}). The in-depth analysis of such additional phenomena is left for future work.


\subsection{Impact on the total luminosity}
\label{subsec: imp_lum}

The presence of outflows might not only lead to MF, but also to the acceleration of non-thermal particles which later deposit energy and momentum into the medium. Therefore the total luminosity of the PBH receives contributions both from the radiation luminosity and the ``non-thermal'' luminosity induced by the outflow, i.e.,
\begin{equation}\label{eq: L_tot}
L_\mathrm{tot} = L_\mathrm{rad} + L_\mathrm{non-th}\,,
\end{equation}
where $L_\mathrm{rad}$ is the same of equation~\eqref{eq: luminosity} and the non-thermal contribution is conveniently parametrized by an additional efficiency parameter~$\epsilon_{\text{non-th}}$ as
\begin{equation}
    L_\text{non-th} = \epsilon_{\text{non-th}} \, \dot{M}_\mathrm{PBH}\,.
\end{equation}

An upper-limit on $\epsilon_{\text{non-th}}$ can be derived assuming that most of the energy released by the accretion gets released as outflows, which upon interaction with the environment (interaction by which MF is realized) converts most of its energy into non-thermal accelerated particles. In such a scenario, $\epsilon_{\text{non-th}}$ would be ultimately determined by the relation between the dominant radiation channel, likely inverse Compton (IC) scattering off CMB photons, non-radiative losses via escape and adiabatic losses. Given the dense CMB photon field at $z\sim 10^2-10^3$, IC could efficiently cool relativistic electrons, with most of the radiation peaking somewhere within the range from soft X-rays to gamma rays. Such a scenario, which would imply $\epsilon_{\text{non-th}}\sim 0.1$, is however highly optimistic, with more reasonable values for $\epsilon_{\text{non-th}}\lesssim 10^{-2}$ (though values $\ll 10^{-2}$ are also possible). A similar scenario has been explored in ref.~\citep{Bosch-Ramon2018} for the case of super-massive BH jets interacting with the medium at high $z$. 

Due to the lack of quantitative predictions based on first principles, a robust (and not highly model-dependent) estimate of $\epsilon_{\text{non-th}}$ has not been obtained yet. Therefore, in order to parametrize its impact on the cosmic medium, we consider three different benchmark values for $\epsilon_{\text{non-th}}$, namely $\epsilon_{\text{non-th}} = 10^{-2},~10^{-4},~10^{-6}$. As a reference, the intermediate case, $\epsilon_{\text{non-th}} = 10^{-4}$, may be interpreted as a scenario where $10$\% of the accretion energy goes to the outflows, of which $10$\% goes to non-thermal particles, which in turn can radiate through IC about 10\% of their energy, 10\% of which in the form of photons with suitable energies (see ref.~\cite{10.1111/j.1365-2966.2012.22029.x} for a discussion in the context of isolated black holes in our galaxy). These benchmark values are compared to the radiation efficiencies obtained for the scenarios in absence of outflows in figure~\ref{fig: plot_eps}. As it can be observed in the figure, the chosen non-thermal emission efficiencies are of the same order of magnitude (depending on the PBH mass) of the radiation efficiencies obtained in the represented cases, hinting to a tight interplay between the two terms contributing to equation~\eqref{eq: L_tot}.

Finally, we note that in this case the gas can be easily ionized through compression by the shock generated by the outflow on large scales. Hence, since the gas cooling time is not small enough for the gas to recombine, the photo-ionization model provides a more realistic description of the physical picture. Nevertheless, we will also consider the collisional ionization model for sake of completeness.


\subsection{Combined effect}
\label{subsec: imp_comb}

\begin{figure}[t]
    \centering
    \includegraphics[width=0.48\textwidth]{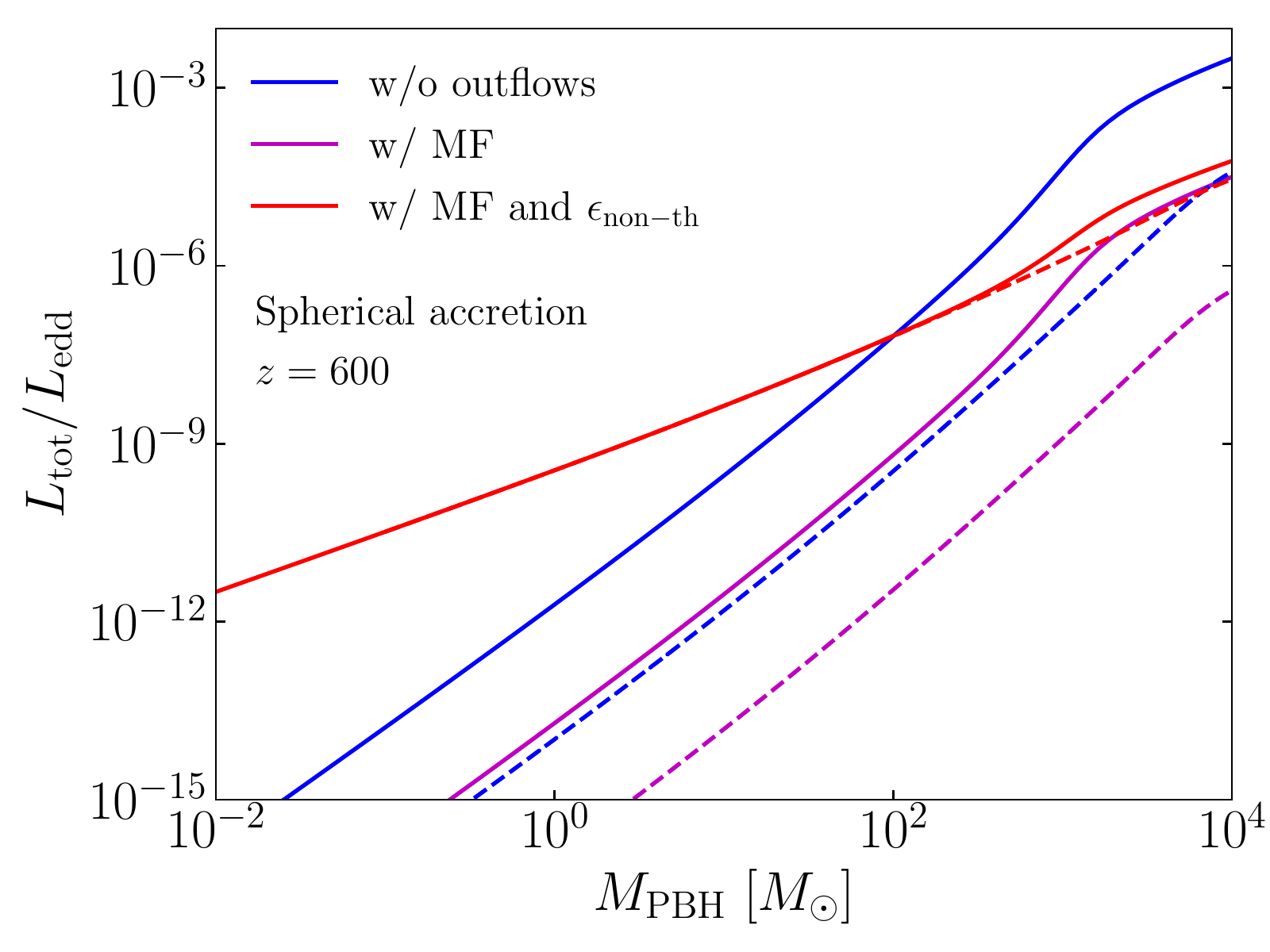}
    \includegraphics[width=0.48\textwidth]{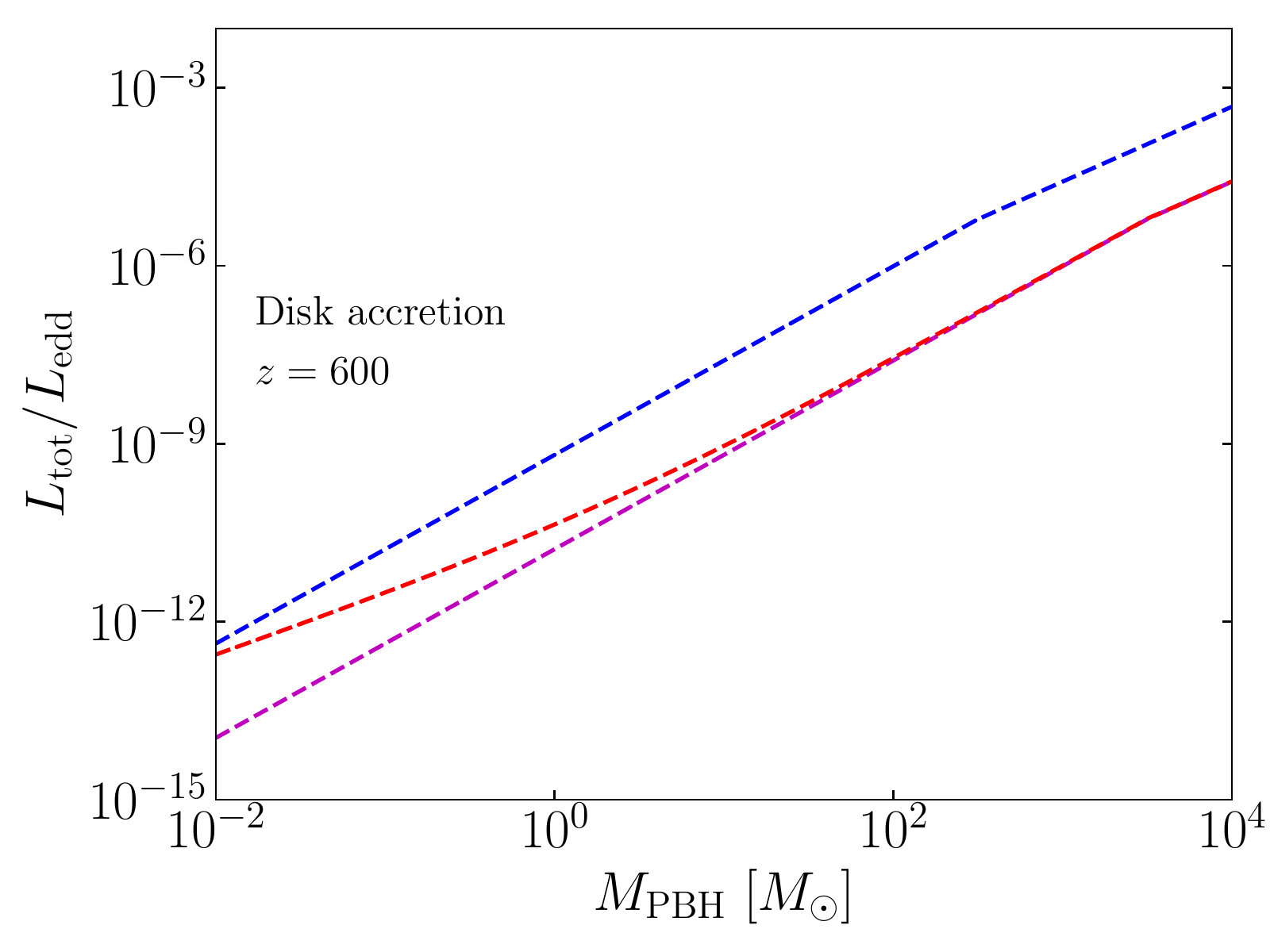}
    \caption{Graphical representation of the impact of the outflow effects (MF and non-thermal emissions) on the total luminosity of the system as a function of the PBH mass. As in figure~\ref{fig: plot_eps}, both the spherical and disk accretion scenarios are shown (on the left and on the right, respectively) as well as the different possible ionization models (photo-ionization in solid and collisional ionization in dashed). In all cases we assume $f_{\rm PBH}=10^{-3}$, $\epsilon_{\rm non-th}=10^{-4}$ and we restrict ourselves only to the representative redshift $z=600$ (as justified in the text).}
    \label{fig: plot_LoverL}
\end{figure}

The impact of MF and non-thermal emissions on the total luminosity is shown in figure~\ref{fig: plot_LoverL} for both accretion geometries (spherical on the left, disk on the right) and ionization models. For sake of succinctness (but with no loss of generality), here we assume~$f_{\rm PBH}=10^{-3}$ and~$\epsilon_{\rm non-th}=10^{-4}$. Furthermore, we also show the results only for the representative case of~$z=600$, which has been shown to be the time around which the CMB anisotropies are most responsive to (almost constant in time) energy injections such as the ones considered here~\cite{Slatyer2015IndirectI}. A more general redshift dependence can be qualitatively inferred from figure~\ref{fig: plot_eps}.

In figure~\ref{fig: plot_LoverL}, the various predictions in absence of outflows are shown in blue and match the expected linear dependence of the radiation luminosity on~$M_{\rm PBH}$ (with different powers depending on the geometry as explained in section~\ref{subsec:geom_acc}) over almost the full PBH mass range. These curves are then suppressed by orders of magnitude when including the role of MF\footnote{That is, since~$L_{\rm rad}\propto \dot{M}_{\rm PBH}^2 \propto f_{\rm LS}^2$ in the spherical accretion case and~$L_{\rm rad}\propto \dot{M}_{\rm PBH}^{1+a} \propto f_{\rm LS}^{1+a}$ in the disk accretion case. By assuming~$f_{\rm LS}=0.1$ (as argued in the previous section) one obtains a suppression of a factor 100 and 40, respectively.}, as represented by the magenta lines. However, this reduction of luminosity can be partially compensated, in particular in the low mass range, when non-thermal emissions are taken into account (red lines).

This implies that when including the effect of outflows there are three possible \text{outcomes}. 
\begin{enumerate}
    \item First of all, if non-thermal emissions are very efficient, their contribution to the total luminosity always dominates over the radiation luminosity and results in an enhanced energy injection with respect to the case in absence of outflows (unless very low $f_{\rm LS}$ values, say $\ll 1$, are considered). This scenario is represented, for instance, in the left panel of figure~\ref{fig: plot_LoverL} in the collisional ionization case. 
    \item On the other hand, the opposite is true if non-thermal emissions are relatively inefficient, as observable in the disk accretion case represented in the right panel of the figure.
    \item Finally, as can be seen from the left panel of the figure for the photoionization case, there might be a range of PBH masses where the total luminosity is enhanced with respect to the case in absence of outflows due to non-thermal emissions (low mass range) and regions where this is instead suppressed because of MF (high mass range).
\end{enumerate}
As discussed in section~\ref{sec:cmb_analysis_constraints}, the balance between these regimes dictates how the final CMB constraints are affected by the presence of outflows.


\section{Impact on thermal history and CMB anisotropies}
\label{sec: ther_hist}

The last ingredient to factor in our analysis is the description of how the emitted radiation impacts the thermal history of the universe. A population of PBHs \textit{injects} energy into the cosmic medium at a rate
\begin{align}\label{eq: E_inj}
    \left.\frac{dE}{dtdV}\right|_{\text{inj}} = \bar{\rho}_{\rm PBH}\, \frac{\left\langle L_\mathrm{tot}\right\rangle}{M_{\rm PBH}} = \bar{\rho}_{\rm cdm}\, f_{\rm PBH}\, \frac{\left\langle L_\mathrm{tot}\right\rangle}{M_{\rm PBH}},
\end{align}
where~$\bar{\rho}_{\rm PBH}$ is the PBH background energy density, $f_{\rm PBH}= \bar{\rho}_{\rm PBH}/\bar{\rho}_{\rm cdm}$ is its fractional form with respect to the DM abundance, and~$\left\langle L_\mathrm{tot}\right\rangle$ is the velocity-averaged total luminosity. However, the amount of energy that is actually \textit{deposited} into the thermal bath can in principle be different, for instance depending on the transparency and energy density of the medium or on the type of emitted particles~\cite{Slatyer2009CMB, Slatyer2013Energy}. Furthermore, different deposition channels have to be accounted for, since the additional energy might not only heat up the medium but also ionize and excite it if the injection happens after recombination~\cite{Chen2004Particle, Padmanabhan2005Detecting, Galli2013Systematic}. Following~\cite{Lucca2019Synergy, Slatyer2015IndirectII, Poulin2017Cosmological}, we express the deposited energy rate in a channel~$c$ as
\begin{align}
    \left.\frac{dE}{dtdV}\right|_{\text{dep}, c} = \left.\frac{dE}{dtdV}\right|_{\text{inj}} f_{\rm eff}\, \chi_c\,,
\end{align}
where~$f_{\rm eff}$ and~$\chi_c$ are the deposition efficiency and deposition fraction for the specific channel, respectively.\footnote{Alternatively, one could have directly calculated the total deposition function~$f_c(z,x_e)$, as in refs.~\cite{Stocker2018Exotic, Liu2019Darkhistory}, but this approach is not currently implemented in the version of CLASS employed here.} In this work we parameterize $f_{\rm eff}$ as in section~IV of ref.~\cite{AliHaimoud2017Cosmic}, while for the calculation of the different deposition fractions we rely on table~V of ref.~\cite{Galli2013Systematic}.

\begin{figure}[t]
    \centering
    \includegraphics[width=0.45\textwidth]{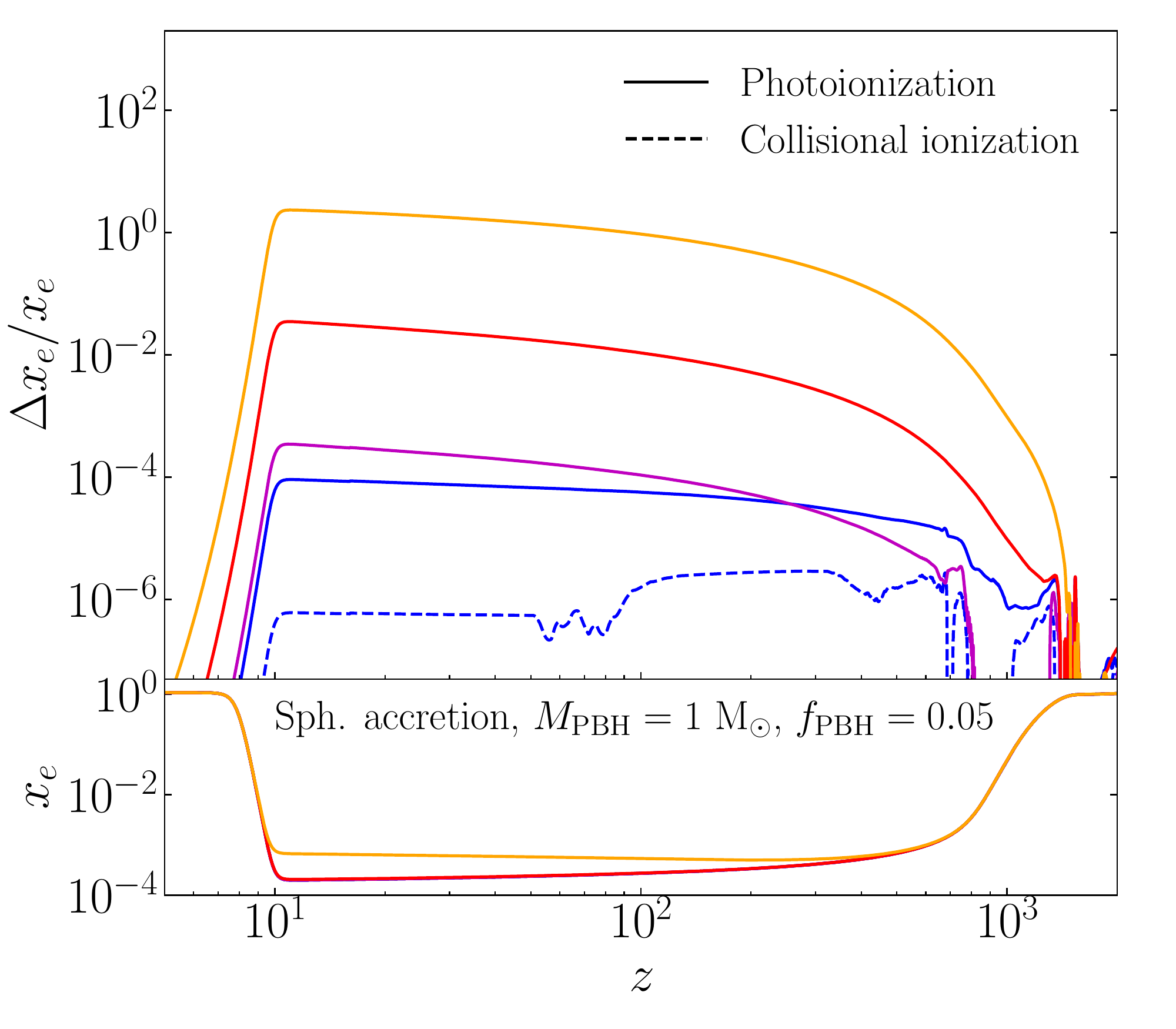}
    \includegraphics[width=0.45\textwidth]{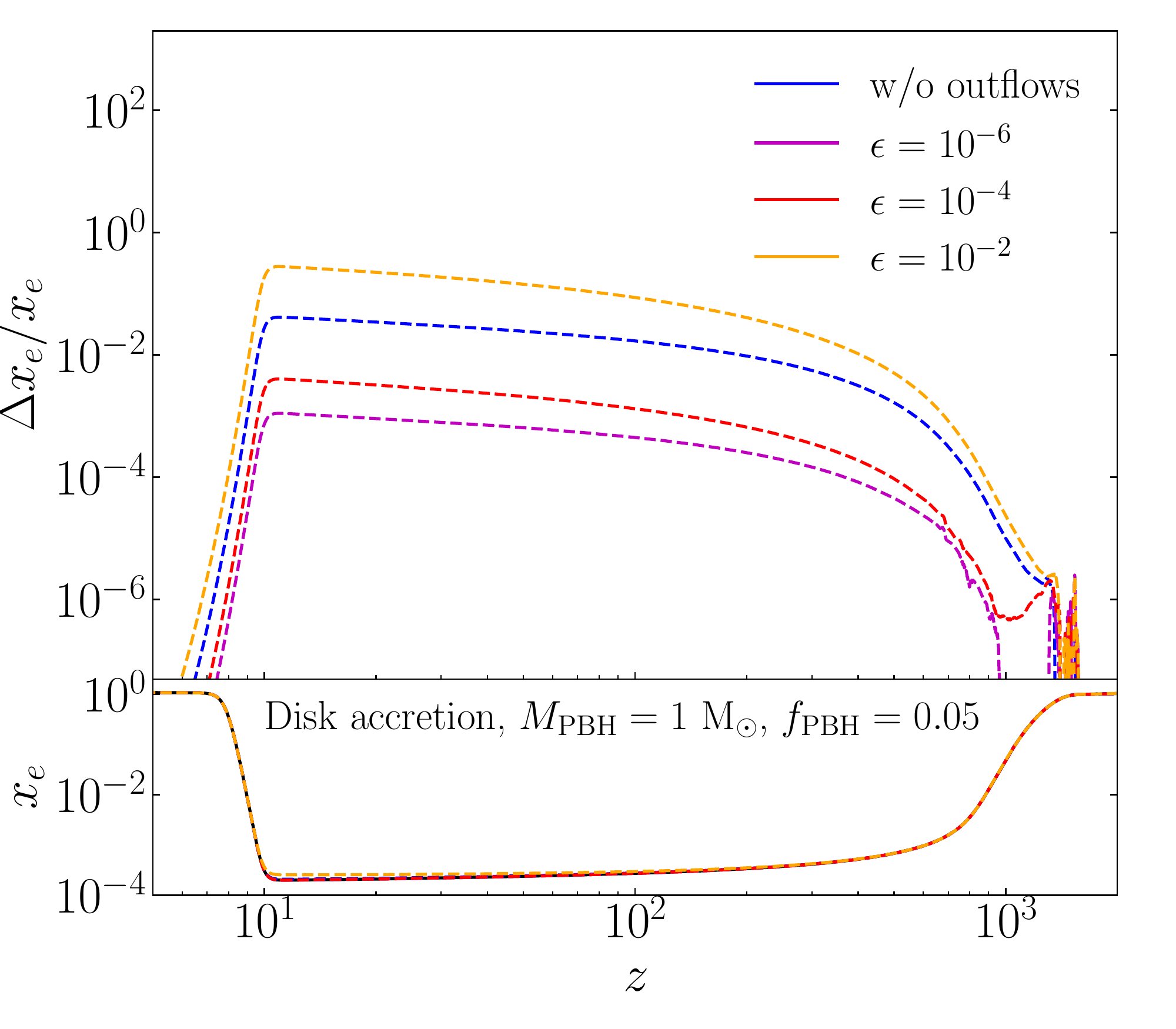}
    \\
    \includegraphics[width=0.45\textwidth]{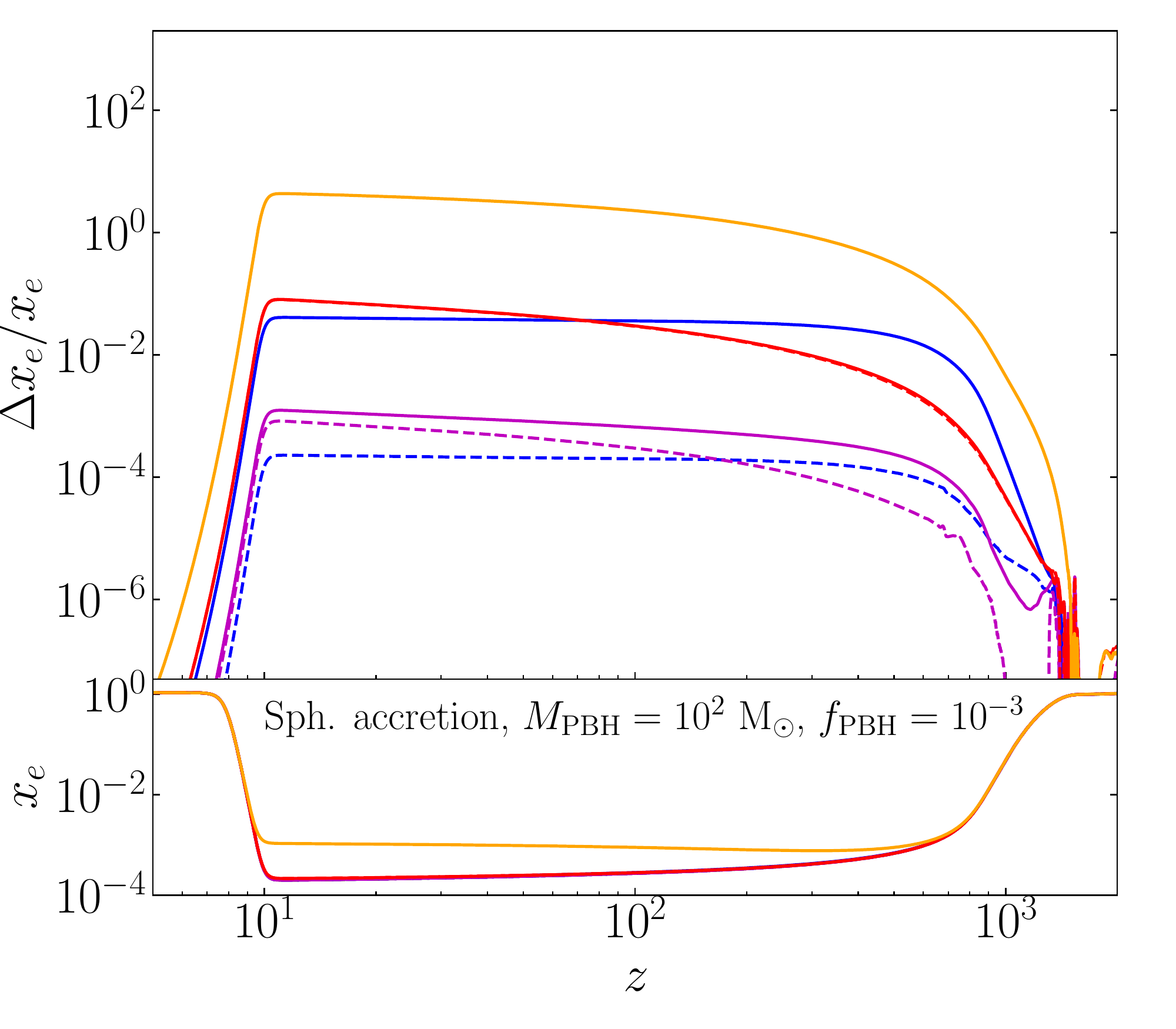}
    \includegraphics[width=0.45\textwidth]{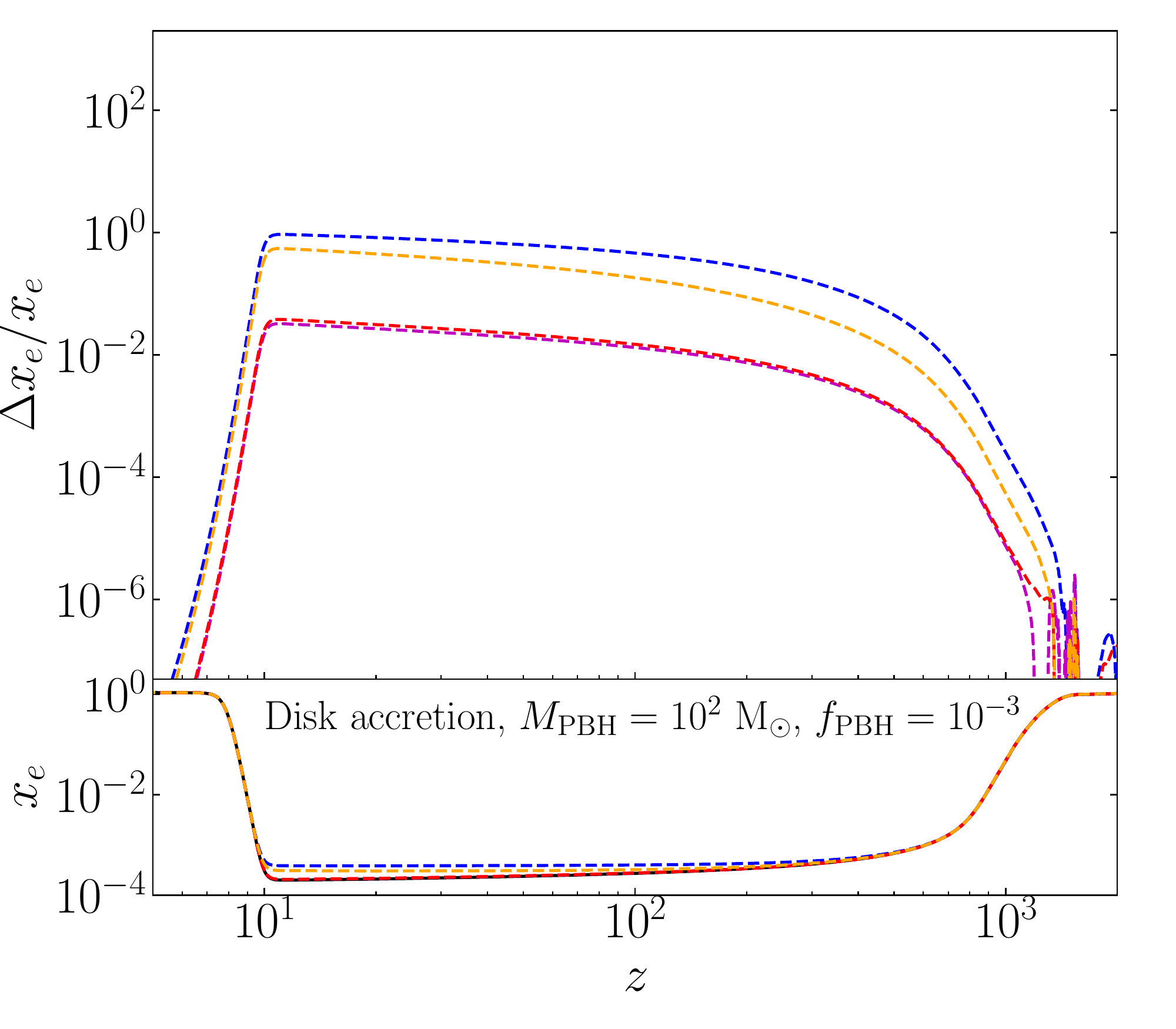}
    \\
    \includegraphics[width=0.45\textwidth]{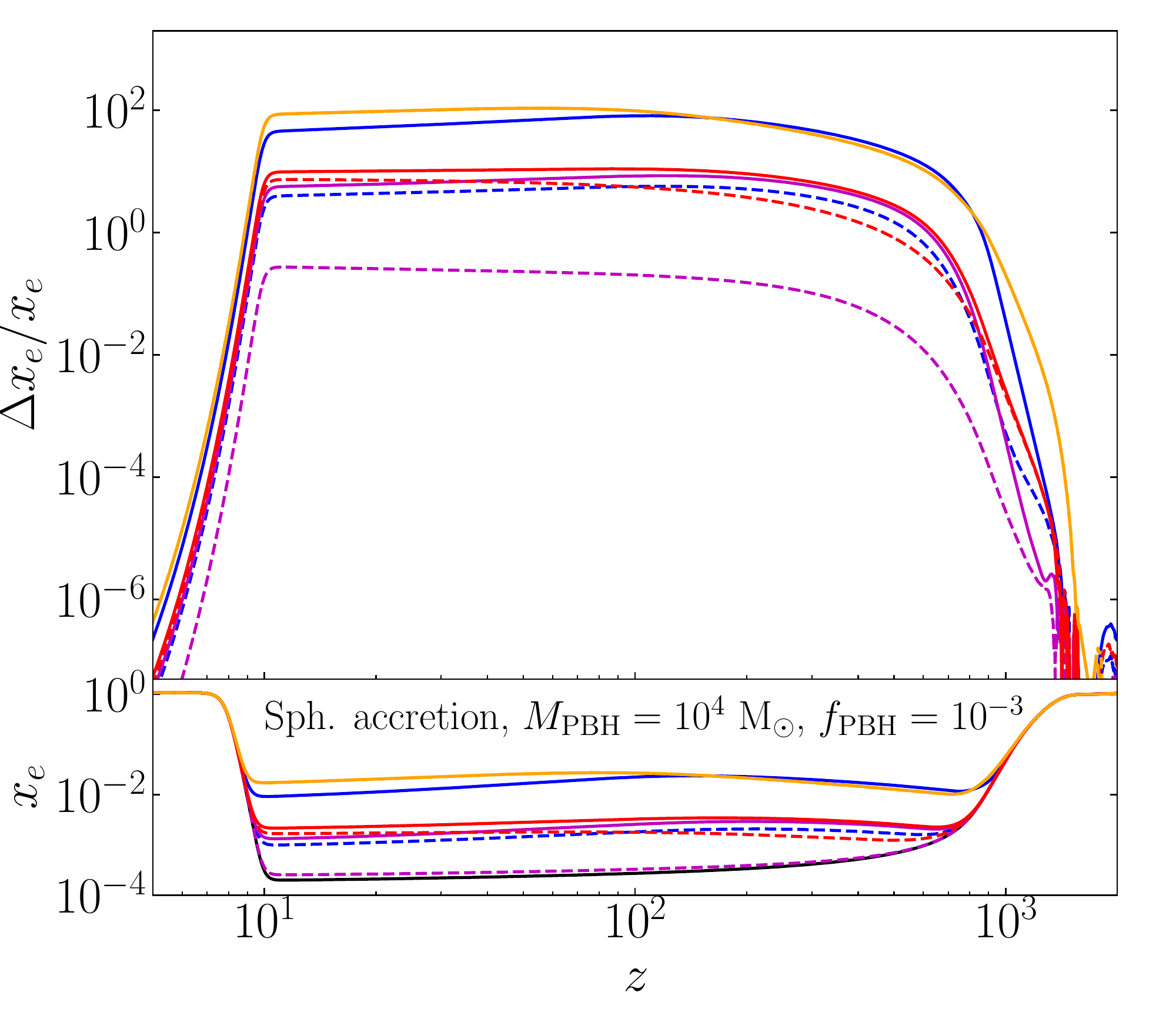}
    \includegraphics[width=0.45\textwidth]{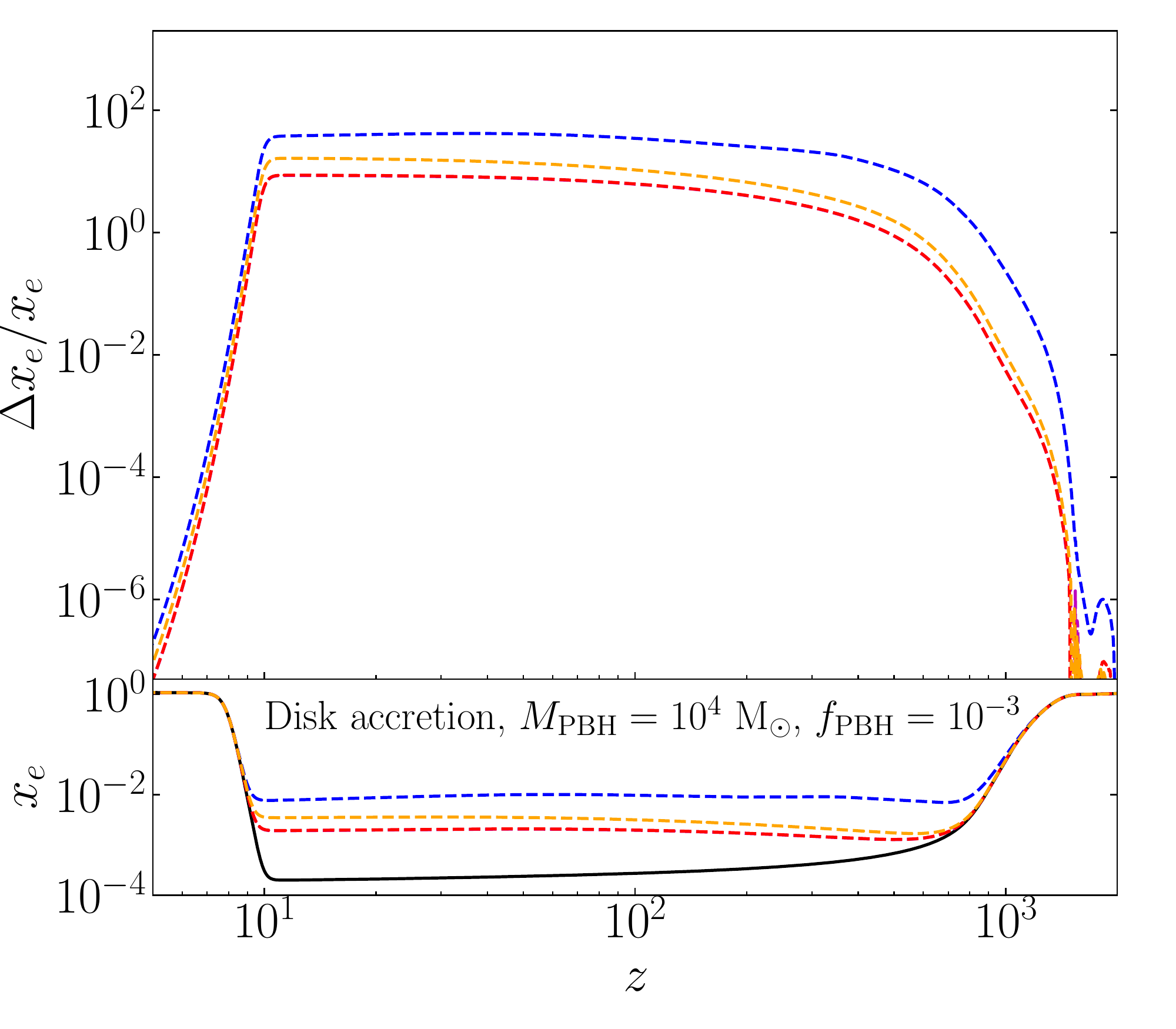}
    \caption{Role of the outflows in the impact on the free electron fraction of the accretion process (shown are the free electron fraction -- lower panels -- and its relative difference with respect to the $\Lambda$CDM prediction -- upper panels). From top to bottom we compare three representative choices of the PBH mass (same as in figure~\ref{fig: plot_eps}), while from left to right we compare the spherical and disk accretion scenarios (for the latter we assumed $\lambda=0.01$ and $\delta=0.1$). In the bottom panels the $\Lambda$CDM prediction is shown in black. The irregular shape of the collisional ionization curve in the top left panel is due to numerical noise.}
    \label{fig: xe_Cl_1}
\end{figure}

\begin{figure}[t]
    \centering
    \includegraphics[width=0.47\textwidth]{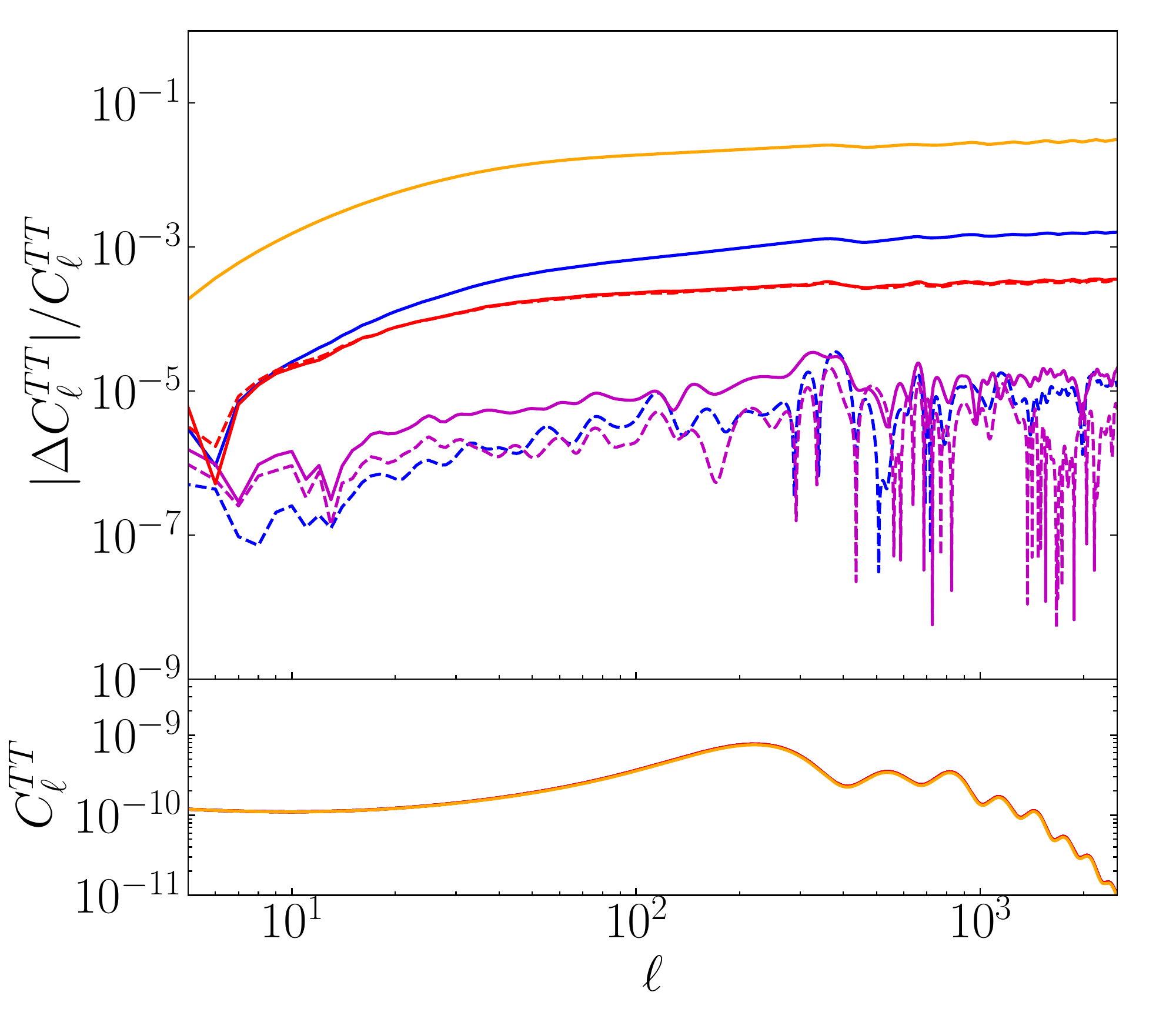}
    \includegraphics[width=0.47\textwidth]{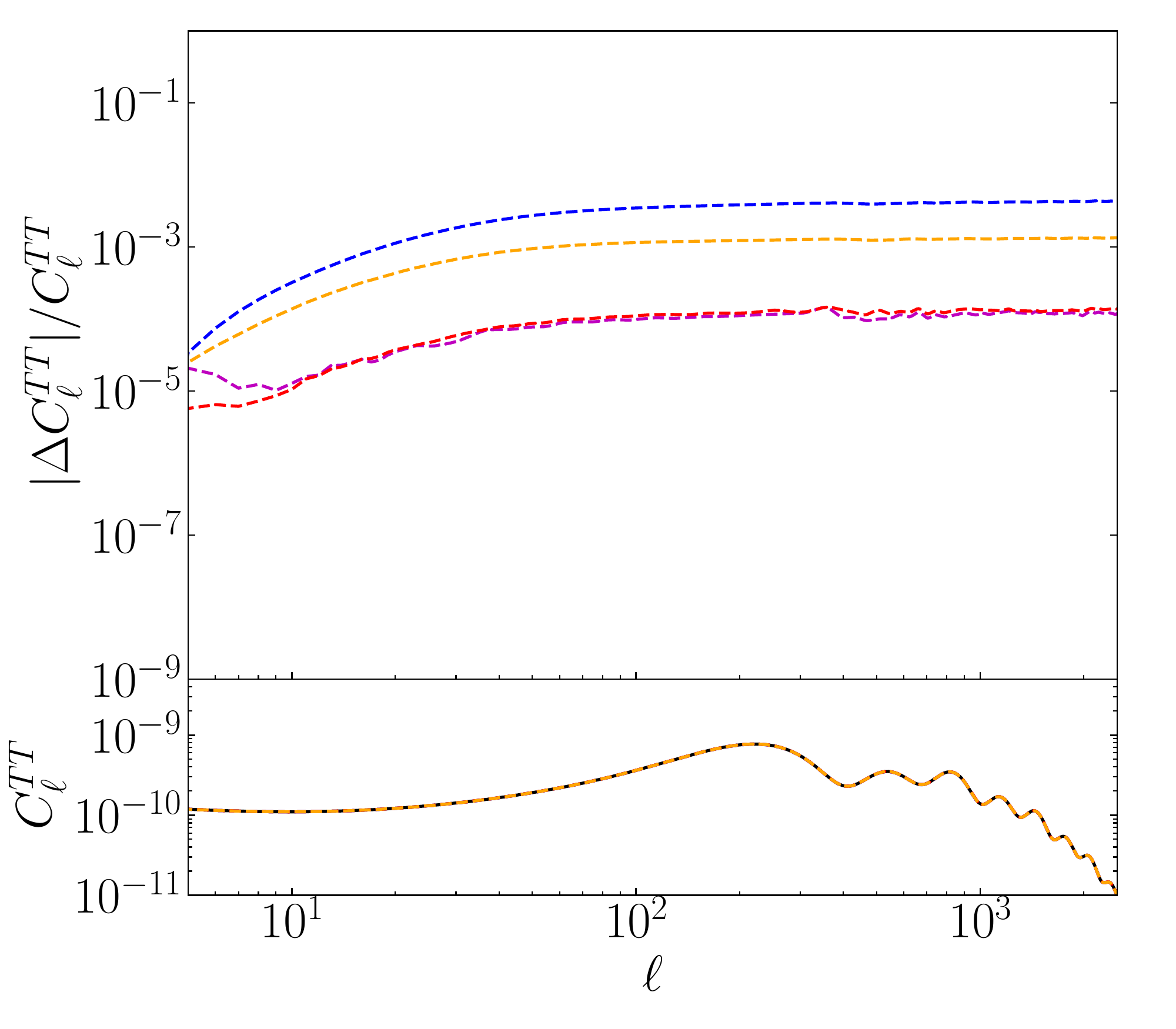}
    \\
    \includegraphics[width=0.47\textwidth]{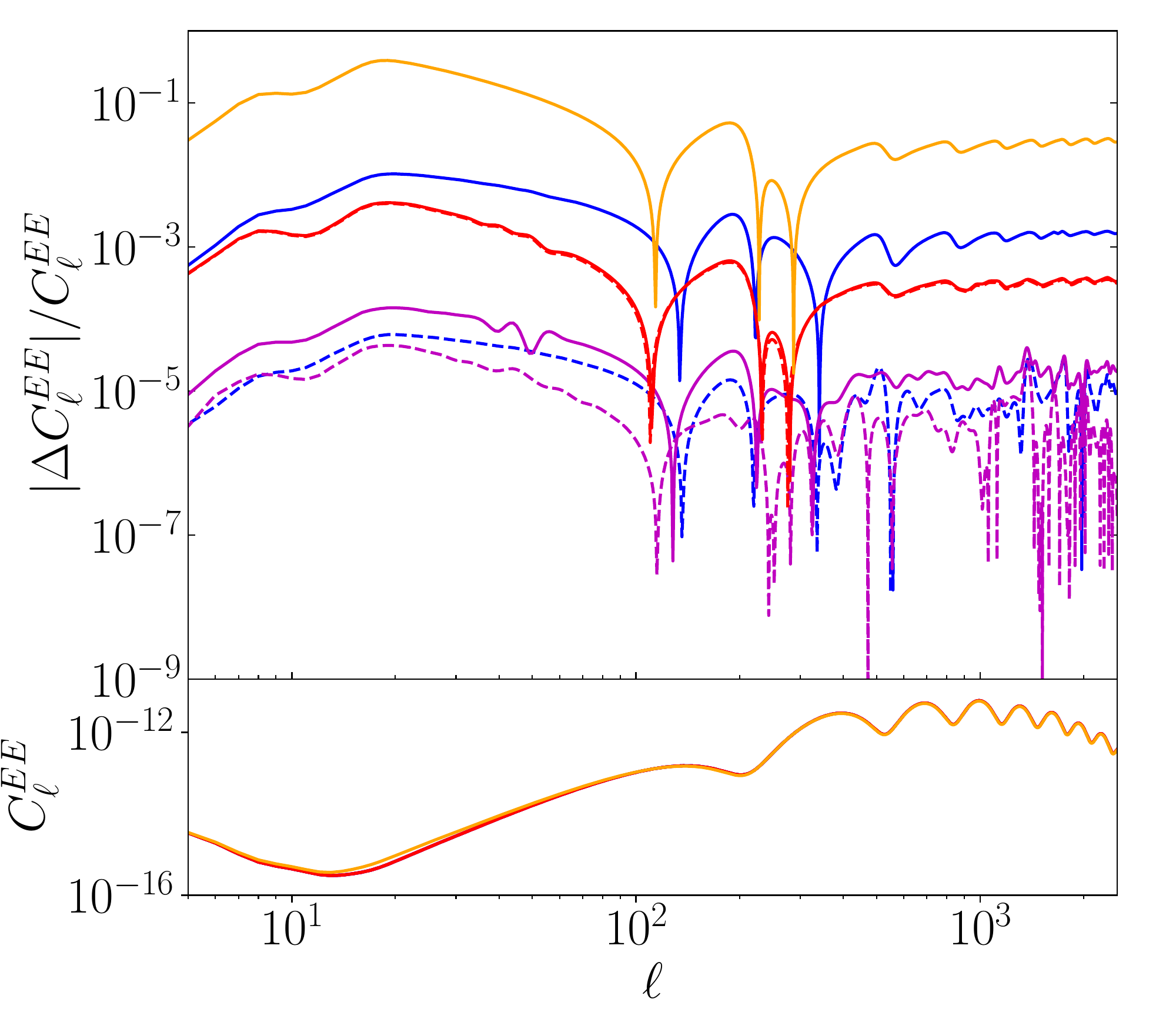}
    \includegraphics[width=0.47\textwidth]{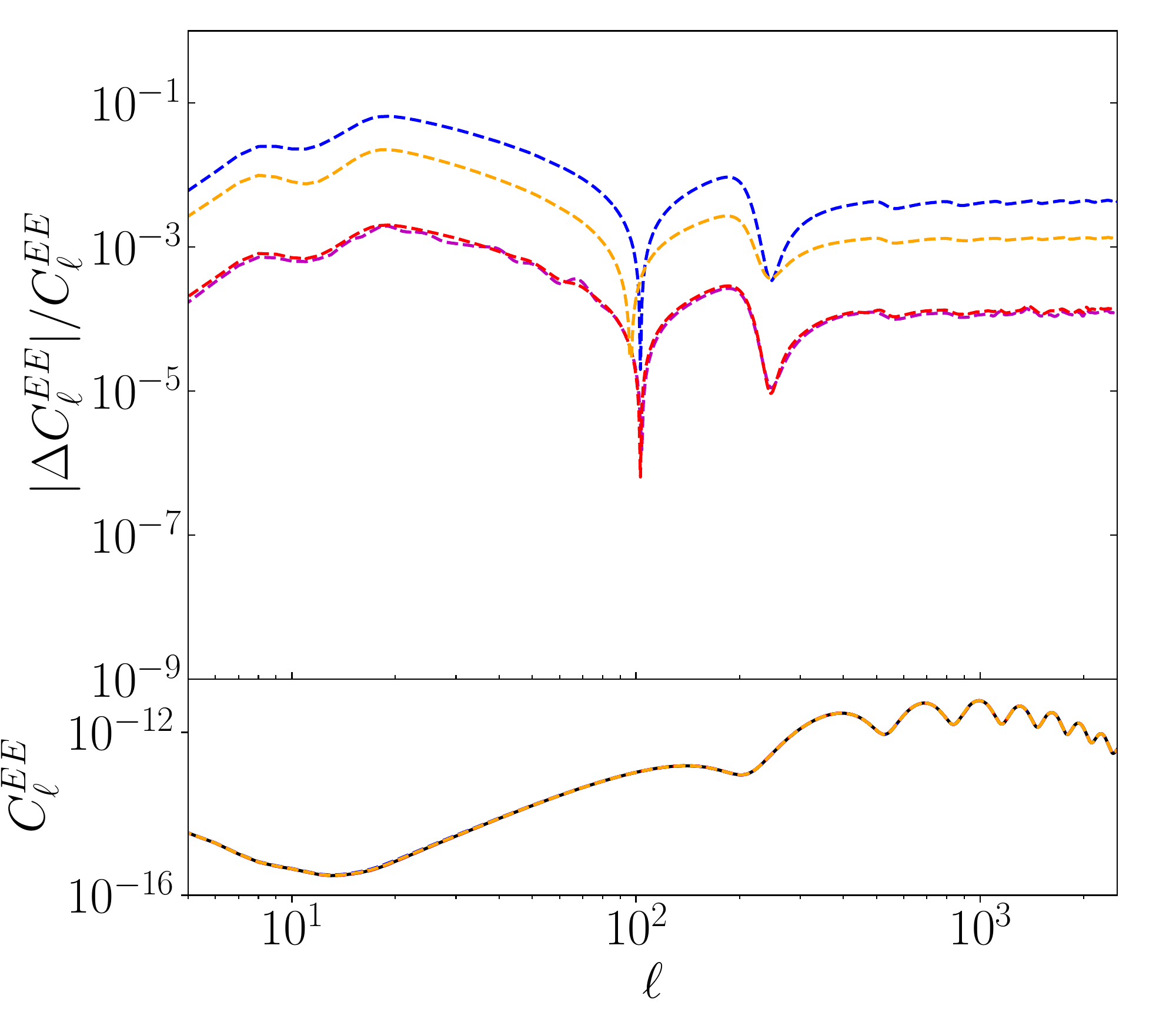}
    \caption{CMB temperature (upper panels) and polarization (lower panels) power spectra for $M_{\rm PBH}=10^2$~$M_\odot$ and $f_{\rm PBH}=10^{-3}$. As in figure~\ref{fig: xe_Cl_1}, the spherical and disk accretion scenarios are shown in the left and right panels, respectively, while solid and dashed lines represent the photo- and collisional ionization cases.}
    \label{fig: xe_Cl_2}
\end{figure}

Once the relation between injected and deposited energy is set, one can analyze how the different accretion scenarios affect the reionization fraction $x_e$, which is ultimately the quantity that determines how much the CMB anisotropy power spectra are affected by the non-standard energy injection. In fact, the main effects of depositing energy into the cosmic medium are twofold: \textit{(i)} delaying the time of last scattering, which induces a slight shift in the position of the peaks in the CMB temperature and polarization power spectra, and \textit{(ii)} effectively increasing the ionization optical depth since the reionization of the universe starts at an earlier time. In figures~\ref{fig: xe_Cl_1} and \ref{fig: xe_Cl_2} we highlight this behavior by comparing the effect on the free electron fraction $x_e$ and the CMB anisotropy power spectra (computed as described in section~\ref{subsec: num}) of the different accretion models considered in this work, with and without outflows, for three representative choices of PBH mass.

From the figures, several conclusions can be drawn. On the left panels of figure~\ref{fig: xe_Cl_1}, which represent the spherical accretion case, we compare the cases with and without outflows and we find that for low PBH masses (upper panel) the contribution from the outflows always dominates over the radiation luminosity. In fact, once MF is accounted for, the mass accretion rate is suppressed by a factor 10 (assuming the fiducial values introduced in section~\ref{subsec: imp_acc}), and the corresponding radiation efficiency becomes significantly lower than any of the cases with a non-zero $\epsilon_{\rm non-th}$. Interestingly, this is true for both ionization models discussed in section~\ref{subsec:ion}, so that no difference is to be seen between the collisional (dashed lines) and photo-ionization (solid lines) cases when including outflows in the low mass range. The situation slightly changes, however, for PBH masses of the order of $10^2$~$M_\odot$ (middle panel) and non-thermal efficiencies lower than $10^{-4}$, where the radiation luminosity in the photo-ionization case becomes comparable to that of the non-thermal outflows after the inclusion of MF and differences between the various ionization processes start to emerge. This behavior is further enhanced for larger PBH masses (bottom panel), although the non-thermal luminosity still dominates for $\epsilon_{\rm non-th}\sim 10^{-2}$.

On the other hand, because of the higher radiation efficiencies reached in the disk accretion scenario (see figure~\ref{fig: plot_eps}), already at relatively low PBH masses (upper panel) the intrinsic radiation efficiency of the accretion is comparable to the non-thermal contribution once MF is taken into account for $\epsilon_{\rm non-th}\sim10^{-6}$. For higher PBH masses (middle and bottom panels) the role of the non-thermal emission becomes negligible and the curves for $\epsilon_{\rm non-th}<10^{-4}$ perfectly overlap, with only a minor difference in the $\epsilon_{\rm non-th}=10^{-2}$ case.

In summary, we expect the contribution from the outflows to be largely determining the luminosity of the system for low masses in the spherical accretion scenario, greatly reducing (if not erasing) the differences between the various ionization models in this regime, while for larger PBH masses the interplay between radiation luminosity, MF and non-thermal emissions starts to become apparent. This differs from the disk accretion case where the intrinsic radiation luminosity of the system is always relevant, and eventually dominates already for PBH masses of the order of $10^2$ $M_\odot$.

In figure~\ref{fig: xe_Cl_2} we also show the impact that these modified thermal histories have on the CMB temperature and polarization anisotropy power spectra. For sake of succinctness, here we focus only on the $M_{\rm PBH}=10^2$ $M_\odot$ case, which we believe already suffices to convey the main message of the section. In particular, for both accretion geometries, we notice the same level of interplay between radiation efficiency, MF and non-thermal effects as in the corresponding panel of figure~\ref{fig: xe_Cl_1}. Interestingly, the contribution of the outflows (visible for instance in the $\epsilon_{\rm non-th}=10^{-2}$ case of the left panels) resembles very closely the smooth behavior already observed in ref.~\cite{Poulin2017CMB} for the disk accretion scenario (see figure 3 of the reference), while the purely spherical accretion case imprints a more oscillatory behavior in the residuals (see, e.g., figure 13 of ref.~\cite{AliHaimoud2017Cosmic}). This might mean, for instance, that a potential observation of a similar signal might be unable to disentangle a spherical accretion scenario dominated by outflows from the disk accretion scenario, regardless of the impact of outflows.


\section{Numerical setup}
\label{subsec: num}

All the relevant cosmological quantities for this work, such as the CMB power spectra, are computed employing the latest version of the Boltzmann solver \textsc{CLASS}~\cite{Lesgourgues2011CosmicI, Blas2011Cosmic}. In particular, we make use of the energy injection treatment discussed in ref.~\cite{Lucca2019Synergy}, which is largely based on the \textsc{ExoCLASS} extension of \textsc{CLASS}~\cite{Stocker2018Exotic}. Recombination is solved using the \textsc{HYREC}~\cite{AliHaimoud2010HyRec,Lee:2020obi} implementation of \textsc{CLASS}, while the energy injection and deposition process is computed using the prescriptions detailed in section~\ref{sec: ther_hist}.

In terms of PBH accretion, we further improve upon the implementation of ref.~\cite{Lucca2019Synergy} (and \cite{Stocker2018Exotic}) by including also the photo-ionization option for the spherical accretion case as well as the calculation for the average of the PBH luminosity over the PBH proper velocities, which we employ as default option for our analysis (see section~\ref{subsec:vel} for further details). The impact of the outflows is implemented as an extension of the spherical and disk accretion cases, where we simply modify the definitions of the accretion rate~$\dot{M}_\mathrm{PBH}$ and of the total luminosity~$L_\mathrm{tot}$, as illustrated by equations~\eqref{eq:mdot_mechanical_feedback} and~\eqref{eq: L_tot}, respectively. When considering disk accretion, we fix $\lambda=0.01$ and $\delta=0.1$, the benchmark values suggested in ref.~\cite{Poulin2017CMB}.

The constraints on the cosmological parameters discussed in section~\ref{sec:cmb_analysis_constraints} for the different accretion models are obtained using the parameter inference code MontePython~\cite{Audren2013Conservative, Brinckmann2018MontePython}. As commonly done in the literature, we perform a number of Markov Chain Monte Carlo (MCMC) scans of the parameter space for fixed values of the PBH mass, which reduces the problem to a 6+1 extension of the $\Lambda$CDM model with
\begin{align}
    \{\omega_b\,, \omega_{\rm cdm}\,, h\,, A_s\,, n_s\,, \tau_{\rm reio} \}+f_{\rm PBH}\,,
\end{align}
where $\omega_b$ and $\omega_{\rm cdm}$ are the baryon and DM physical energy densities, respectively, $h$ is the dimensionless Hubble rate, $A_s$ and $n_s$ are the amplitude and scalar spectral index of the primordial power spectrum, and $\tau_{\rm reio}$ is the reionization optical depth.\footnote{In figures~\ref{fig: v_rel}-\ref{fig: xe_Cl_2} we always assume for these parameters the mean values reported in table~2 of ref.~\cite{Aghanim2018PlanckVI} for the Planck+BAO combination.} In the scenarios of interest, PBHs are described by the abundance parameter~$f_\mathrm{PBH}=\bar{\rho}_\mathrm{PBH}/\bar{\rho}_\mathrm{cdm}$, i.e., by the fraction of DM in form of PBHs. We consider temperature, polarization and lensing information from the Planck 2018 mission~\cite{Aghanim2018PlanckVI} (explicitly, we use the high-$\ell$ TTTEEE, low-$\ell$ EE, low-$\ell$ TT and lensing likelihoods) and determine the MCMCs to be converged with the Gelman-Rubin criterium $|R-1|<0.02$~\cite{Gelman1992Inference}. 


\section{CMB constraints on PBH abundance}
\label{sec:cmb_analysis_constraints}

In this section we derive the corresponding CMB anisotropy constraints on the PBH abundance for both accretion geometries and ionization models described in section~\ref{sec: state_art} and focusing on the impact of outflows for different choices of the non-thermal emission efficiency as explained in section~\ref{sec:mechanical_feedback}. To ease the comparison with the literature, first of all we derive the aforementioned constraints in the context of a monochromatic PBH mass distribution in section~\ref{subsec:constraints_monochromatic}. Then, for sake of generality, we also recast our constraints in terms of a popular choice of extended mass distribution in section~\ref{subsec:constraints_extended}. On the basis of these results, in section~\ref{subsec:implications} we also comment on their implications for the LVK window.


\subsection{Constraints for monochromatic mass distributions}
\label{subsec:constraints_monochromatic}

We begin by assuming that PBHs have a monochromatic mass distribution (MMD), i.e., all PBHs have the same mass. In full generality, a PBH mass distribution is described by the fractional abundance function~\cite{Bellomo:2017zsr}
\begin{equation}
    \frac{df_\mathrm{PBH}}{dM_\mathrm{PBH}} = f_\mathrm{PBH} \frac{d\Phi_\mathrm{PBH}}{dM_\mathrm{PBH}},
\end{equation}
where~$d\Phi_\mathrm{PBH}/dM_\mathrm{PBH}$ describes the shape of the PBH mass distribution, it is normalized to unity by construction and in the monochromatic case it reads as
\begin{equation}
    \frac{d\Phi_\mathrm{PBH}}{dM_\mathrm{PBH}} = \delta^D(M_\mathrm{PBH}-M^\star_\mathrm{PBH}).
\end{equation}
In this work we choose~$M^\star_\mathrm{PBH}$ such that it scans the mass range~$[10^{-2},10^4]\ M_\odot$, computing for each PBH mass the respective upper limit on the PBH abundance.

\begin{figure}[t]
    \centering
    \includegraphics[width=0.47\textwidth]{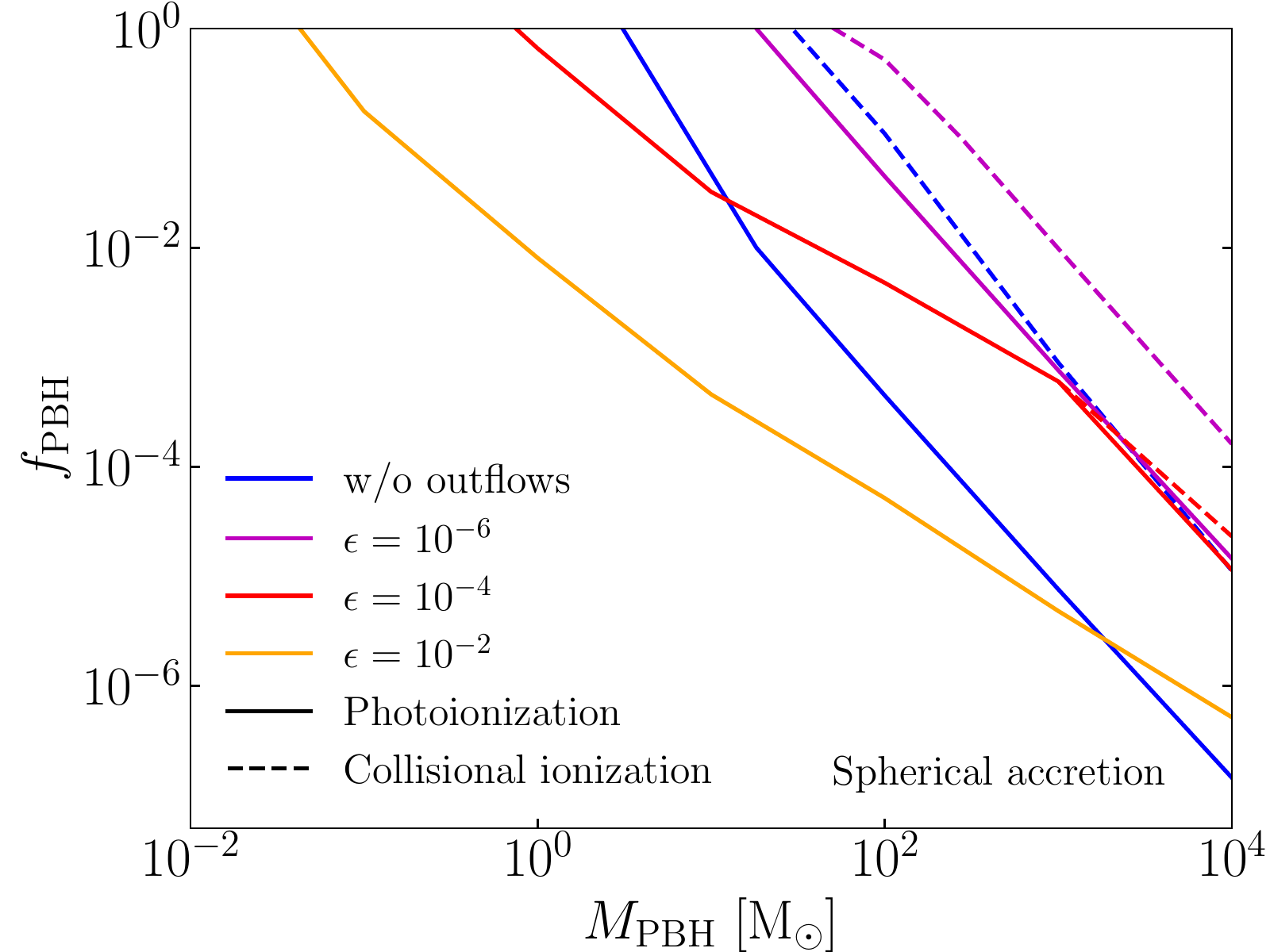}
    \includegraphics[width=0.47\textwidth]{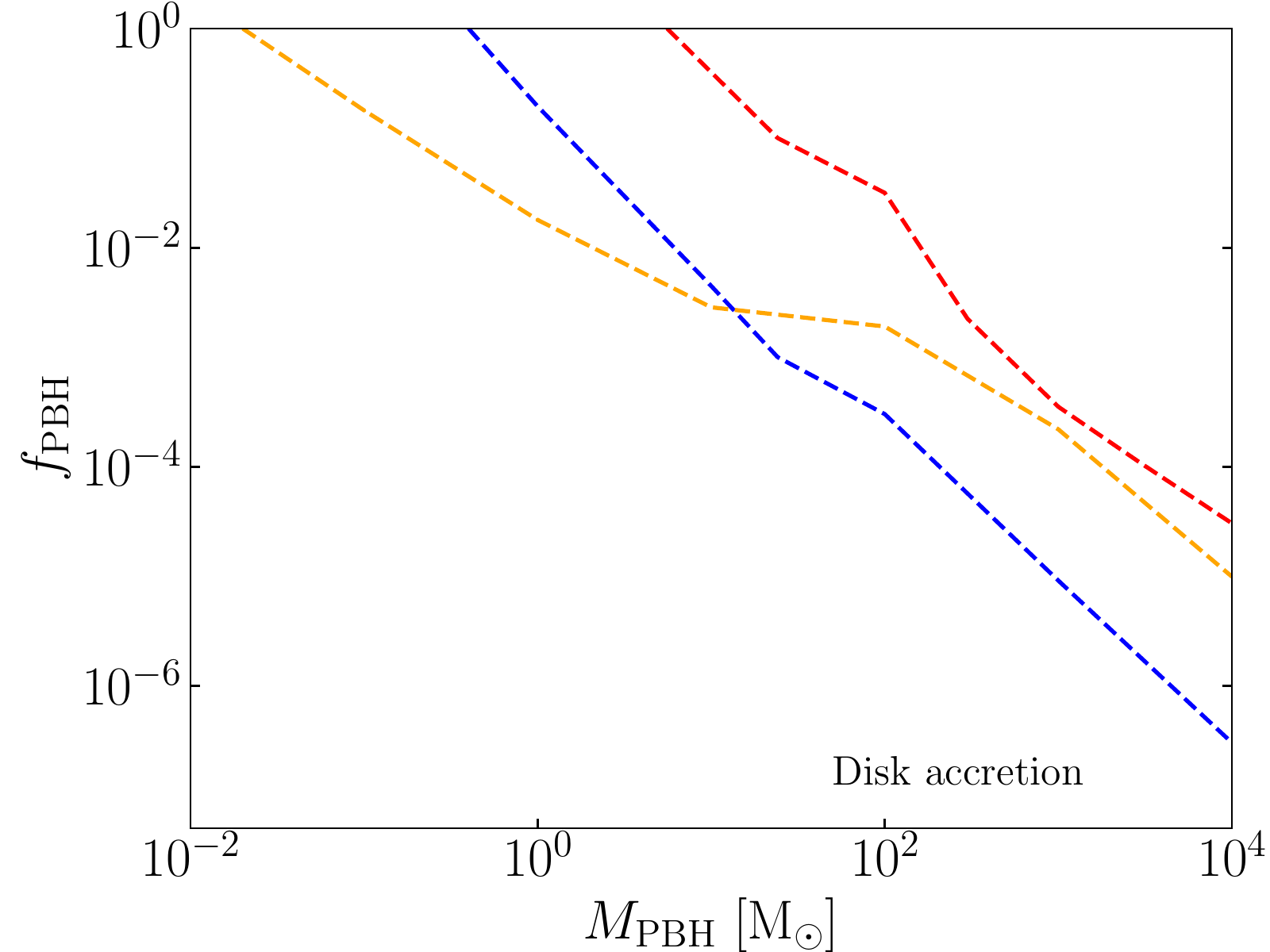}
    \caption{Impact of outflows on the CMB constraints on the fractional PBH abundance for the two main types of accretion geometry (spherical on the left and disk on the right) and ionization models (photo-ionization in solid and collisional ionization in dashed) assuming a MMD. The blue contours represent the scenarios without outflows and update the bounds first derived in \cite{AliHaimoud2017Cosmic} (left) and \cite{Poulin2017CMB} (right), while the magenta, red and organge lines assume MF with $f_{\rm LS}=0.1$ and non-thermal emissions with $\epsilon_{\rm non-th}=10^{-6,-4,-2}$, respectively.}
    \label{fig: bounds}
\end{figure}

We show in figure~\ref{fig: bounds} the corresponding~$95\%$ CL upper bounds on the fractional PBH abundance $f_{\rm PBH}$ for different accretion mechanisms, ionization choices and outflows scenarios. In particular, in each of the two subplots, which display the bounds for the spherical accretion case on the left and the disk accretion case on the right, we report the upper bounds on $f_{\rm PBH}$ for the two aforementioned ionization models (photo-ionization reported as solid lines and collisional ionization as dashed lines) as well as for different choices of the parameters describing the outflow modelling. As in the previous sections, we have the scenario without any outflow contribution in blue, while the contours including MF with the benchmarking value of $f_{\rm LS}=0.1$ and non-thermal emissions are reported in magenta, red and orange for $\epsilon_{\rm non-th}=10^{-6,-4,-2}$, respectively.

Focusing first on the left panel of figure~\ref{fig: bounds}, i.e., on spherical accretion, we observe that the constraints obtained in absence of outflows are approximately one order of magnitude more stringent than those reported in ref.~\cite{AliHaimoud2017Cosmic} for both ionization models. We attribute this discrepancy to the difference in methodology\footnote{In ref.~\cite{AliHaimoud2017Cosmic}, for instance, an approach based on a Fisher-information matrix has been employed, while here we make use of an MCMC analysis.} and Planck data release employed here compared to the one in ref.~\cite{AliHaimoud2017Cosmic}. The bounds for the collisional ionization case are in fact broadly consistent with those presented in ref.~\cite{Serpico:2020ehh}, where a more similar analysis was conducted. In terms of impact of the outflows, for low non-thermal efficiencies the role of MF alone is dominant, leading to an overall suppression of the constraints by approximately $1-2$ orders of magnitude, as explained in section~\ref{subsec: imp_comb}. For intermediate values of~$\epsilon_{\rm non-th}$ the interplay between MF and non-thermal emissions becomes more balanced, with the latter most predominately contributing to the total luminosity until PBH masses of the order of~$10^3$~$M_\odot$, above which the radiation luminosity dominates despite the presence of MF. Nevertheless, also in this scenario we observe a suppression of the bounds with respect to the case without outflows, with only a minor enhancement for masses below~$10$~$M_\odot$. The situation changes for very efficient non-thermal outflows, which dominate the accretion emission over the whole PBH mass range, thereby arasing the difference between photo- and collisional ionization (which perfectly overlap in the figure), and significantly strengthen the constraints down to PBH masses of the order of 0.1~$M_\odot$ for~$f_{\rm PBH}=1$.

On the other hand, since in the context of disk accretion the intrinsic radiation luminosity is much higher than in the spherical accretion case, in the right panel of figure~\ref{fig: bounds} we notice that even for non-thermal efficiencies as high as~$10^{-4}$ the only relevant impact of outflows is MF, which suppresses the constraints by about $1-2$ orders of magnitude. For this reason, the contours for the $\epsilon_{\rm non-th}=10^{-4}$ and $\epsilon_{\rm non-th}=10^{-6}$ cases perfectly overlap in the right panel of the figure. Only for~$\epsilon_{\rm non-th}$ of the order of~$10^{-2}$ non-thermal emissions start to play a significant role, enhancing the constraints for PBH masses below 10~$M_\odot$ but still leading to a suppression thereof for larger PBH masses.

Focusing on the shapes of the constraints, one expects the behaviour of the bounds to follow the dependence of the deposited energy on the free parameters of the model, i.e., $f_{\rm PBH}$ and~$M_{\rm PBH}$, which roughly boils down to $\text{d}E/\text{d}t\text{d}V|_{\rm dep} \propto f_{\rm PBH} M_{\rm PBH}\lambda (\epsilon_{\rm non-th}+M_{\rm PBH}^2\lambda$). In the case of spherical accretion the values of~$\lambda$ at~$z\sim\mathcal{O}(10-10^3)$ increase as a function of the mass (see figure 4 of ref.~\cite{AliHaimoud2017Cosmic}), which explains why the relation between PBH abundance and mass is not linear, with the bounds becoming comparatively less stringent the lower the PBH mass (a similar behaviour is also to be observed in figure~\ref{fig: plot_LoverL}). This relation, and in particular the different dependence on the PBH mass, also explains the difference in slope of the $f_{\rm PBH}-M_{\rm PBH}$ upper bounds between the radiation and non-thermal emission dominated regimes. In the disk accretion scenario the value of~$\lambda$ is fixed to a fiducial value, and therefore the upper bounds approach a linear dependence more closely than in the spherical case. Nevertheless, as it becomes clear in particular in the~$\epsilon_{\rm non-th}=10^{-4,-6}$ cases, the parametric form of~$\epsilon$ used in ref.~\cite{Xie:2012rs} still introduces a non-trivial dependence of the constraints on the PBH mass.


\subsection{Constraints for extended mass distributions}
\label{subsec:constraints_extended}

Even if for practical purposes it is more convenient to obtain constraints for MMDs, in reality it is well known that PBH populations would likely have an extended mass distribution (EMD). Different PBH formation mechanisms are generally responsible for different EMDs, although we can identify two popular benchmark classes of EMD: \textit{power-law} and \textit{lognormal}. Broadly speaking, the first class is typically associated with the collapse of large density perturbations or cosmic strings, see e.g., refs.~\cite{carr:pbhfrominhomogeneities, hawking:pbhfromstrings}, while the latter is connected to the presence of large peaks in the primordial power spectrum, see e.g., refs.~\cite{ivanov:pbhsfrominflation, bellido:pbhsfrominflation, ivanov:pbhsfrominflationII}. 

Recent works, as for instance  refs.~\cite{leach:runningmassmodel, drees:runningmassmodel, drees:runningmassmodelII, kawasaki:axioncurvatonmodel, kohri:axioncurvatonmodel, bellido:inflectionpointmodel, germani:inflectionpointmodel, kannike:doubleinflationmodel, motohashi:slowrollbreaking, ballesteros:loopcorrectionsmodel, ozsoy:stringtheorymodel, cicoli:stringtheorymodel, dalianis:alphaattractorsmodel}, have shown significant interest in models where PBHs are generated from peaks in the primordial curvature power spectrum. Such effort has been mainly driven by the intrinsic connection between the primordial power spectrum and inflationary dynamics, which is known to be compatible with the single-field slow-roll scenario only during~$6/8$ e-folds of the at least~$50$ required to solve the horizon and flatness problems~\cite{ade:planckinflation2013,Ade2015PlanckXX,akrami:planckinflation2018}. Therefore establishing robust bounds on the PBH abundance, or even just their presence, is a key ingredient in reconstructing the primordial power spectrum at scales of order~$\mathcal{O}(10^5-10^{15})\ \mathrm{Mpc}^{-1}$ currently not accessible by other cosmological observables~\cite{carr:pbhmassspectrum, carr:powerspectrumconstraintsI, carr:powerspectrumconstraintsII, josan:powerspectrumconstraints, cole:powerspectrumconstraints, mifsud:powerspectrumconstraints, satopolito:powerspectrumconstraints, akrami:powerspectrumuncertaintities, kalaja:powerspectrumconstraints, gow:powerspectrumconstraints, Schoeneberg2020Constraining}. For this reason, here we focus on deriving abundance constraints for lognormal EMDs
\begin{equation}
    \frac{d\Phi_\mathrm{PBH}}{dM_\mathrm{PBH}} = \frac{e^{-\frac{\log^2(M_\mathrm{PBH}/\mu)}{2\sigma^2}}}{\sqrt{2\pi}\sigma M_\mathrm{PBH}},
\end{equation}
characterized by two parameters, the mean~$\mu$ and standard deviation~$\sigma$. More realistic mass distribution models would require a greater number of parameters to describe the EMD~\cite{gow:realisticpbhemd}, and therefore in this work we choose the simplest extension of the MMD case. The same procedure can be implemented also in the case of power-law EMDs, as done for instance in ref.~\cite{Bernal:2017nec}.

\begin{figure}[t]
    \includegraphics[width=0.99\columnwidth]{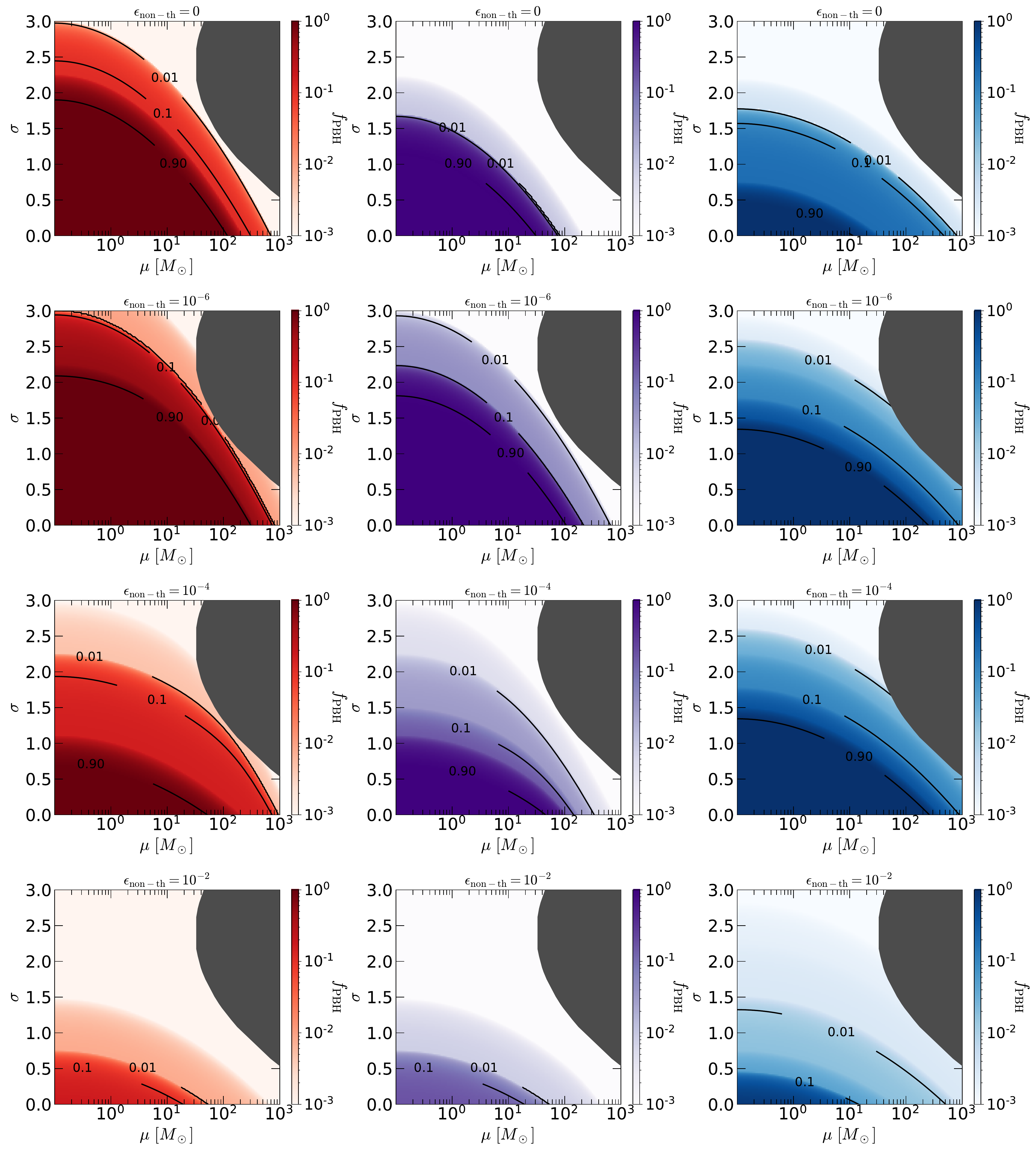}
    \caption{Same as in figure~\ref{fig: bounds}, but for a lognormal EMD. The figure shows the collisional and photoionization regimes for the spherical accretion case (left and center panels, respectively) and the disk accretion model (right panels). We report here all the cases in absence of MF and non-thermal emission (top panels), and with those effects included (middle and bottom panels) for different non-thermal efficiencies. The grey shaded area represents the region of parameter space that cannot be described appropriately by the theoretical model employed in this work.}
    \label{fig:emd_bounds}
\end{figure}

Interpreting abundance constraints obtained for MMDs as abundance constraints for EMDs is not straightforward~\cite{carr:comparison1}, although several methods have been proposed to infer the latter from the former (see e.g., refs.~\cite{Bellomo:2017zsr, carr:comparison1,carr:comparison2}). In this work we follow the approach of ref.~\cite{Bellomo:2017zsr}, based on the concept of \textit{equivalent mass}. In the case of interest, if PBHs have an EMD, equation~\eqref{eq: E_inj} reads as
\begin{equation}
    \left.\frac{dE}{dtdV}\right|_{\text{inj}} = \bar{\rho}_{\rm cdm} f_\mathrm{PBH} \int dM_\mathrm{PBH} \frac{d\Phi_\mathrm{PBH}}{dM_\mathrm{PBH}} \frac{\left\langle L_\mathrm{tot}\right\rangle}{M_{\rm PBH}},
\label{eq:injected_energy_emd}
\end{equation}
showing that it is possible for different EMDs to inject the same amount of energy into the cosmic medium, making it possible for PBH populations in those models to be equally abundant. At a practical level, it is always possible to associate the impact on a cosmological observable of a given EMD, and therefore its abundance~$f^\mathrm{EMD}_\mathrm{PBH}$, to the same-magnitude effect generated by a MMD with equivalent mass~$M_\mathrm{eq}$, i.e., to its abundance~$f^\mathrm{MMD}_\mathrm{PBH}$. Despite its simplicity, this method provides an easy analytical tool to compute constraints for EMDs starting from the MMD analysis of section~\ref{subsec:constraints_monochromatic}. The validity of this method has been already explicitly proved in ref.~\cite{Bernal:2017nec} for the CMB case or, for instance, in ref.~\cite{manshanden:accretionconstraint} for other abundance constraints also based on accretion physics.

In figure~\ref{fig:emd_bounds} we show the PBH abundance constraints for the lognormal EMD for all geometries, ionization models and non-thermal emission efficiencies already employed in the previous section. We report the new conversion formula used for all the cases at hand in appendix~\ref{app:emd_converting_relations}. The gray shaded regions in the plots indicate the values of $\mu$ and $\sigma$ for which the EMDs extend beyond $10^4\, M_{\odot}$ and the theoretical models employed in this work break down as discussed in section~\ref{subsec:add_eff} (see ref.~\cite{Bellomo:2017zsr} for further details). By comparing the cases with only radiative efficiency to the ones with also non-thermal emission, we note that as soon as the MF effect starts to play a role, the allowed parameter space increases significantly with respect to the case of radiative emission only. On the other hand, as soon as the non-thermal emission grows, the parameter space becomes very tightly constrained, as can be seen in the bottom panels. In other words, because of these two competing effects, the same EMD can be either ruled in or ruled out depending on the details of the accretion and emission models.


\subsection{Theoretical uncertainties and implications for the LVK mass range}
\label{subsec:implications}

\begin{figure}[t]
    \centering
    \includegraphics[width=0.65\columnwidth]{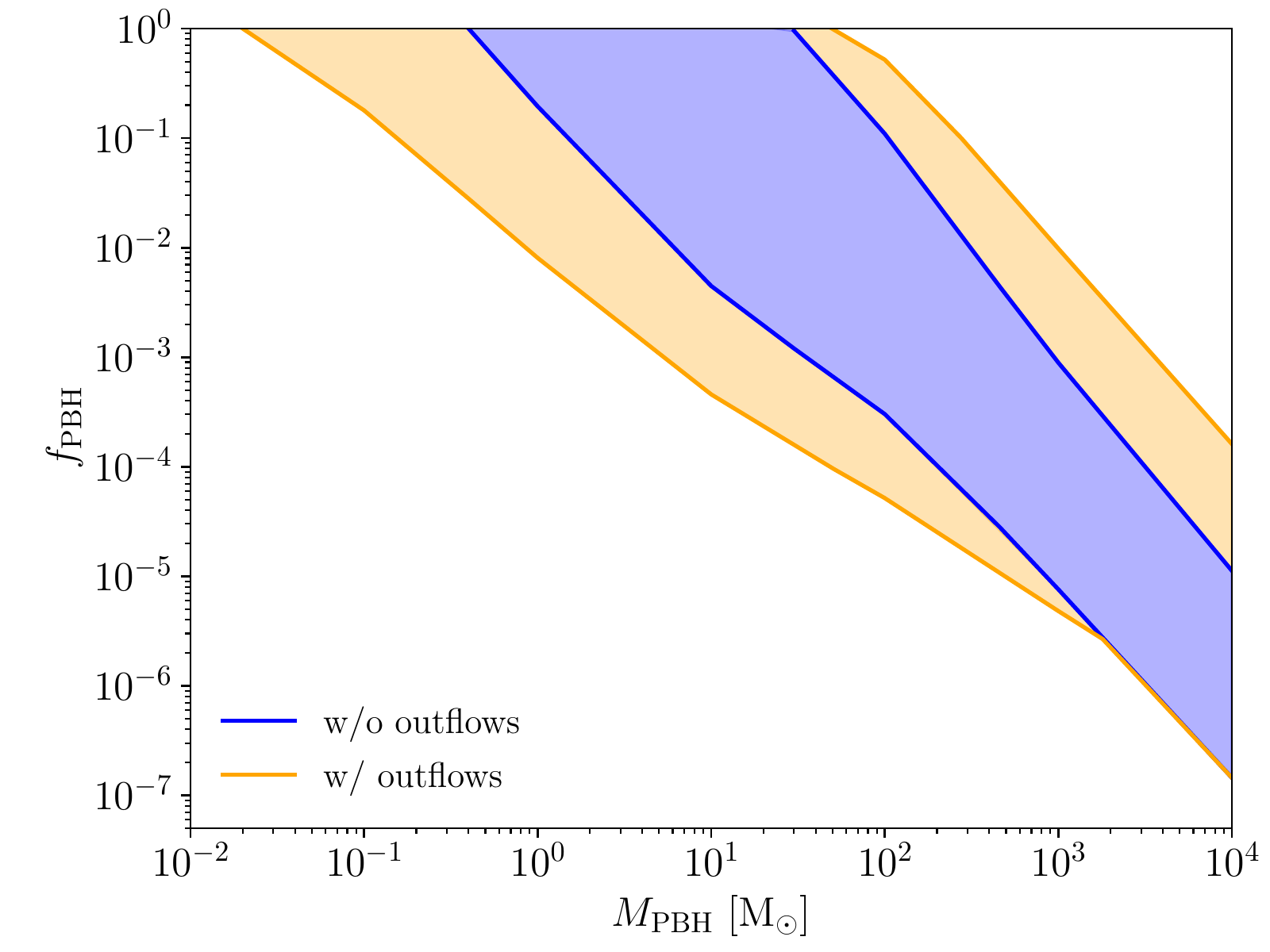}
    \caption{Uncertainty bands of both geometries and ionization models with (orange) and without (blue) the inclusion of outflows. The corresponding filled regions of parameter space represent the region where the true constraint lies.}
    \label{fig: bounds_v2}
\end{figure}

The analysis presented in this work clearly highlights what is one of the underlying issues with existing constraints on PBHs: the unknown magnitude of the theoretical error associated to the modelling of the accretion process. Reducing this uncertainty is crucial given that existing and future GW observatories are and will be sensitive in the~$\mathcal{O}(10-10^4)\ \mathrm{Hz}$ frequency range, i.e., the typical frequency range where the signal of coalescing compact objects with masses~$\mathcal{O}(10-10^3)\ M_\odot$ is expected to be. An accurate estimation of the PBH abundance upper limits would in fact allow us to speculate more realistically about what amount of detected events can have a primordial origin.

Nevertheless, such an accurate description of the accretion process is still not available. Therefore, we present our findings in figure~\ref{fig: bounds_v2} as ``uncertainty bands'' around the true, unknown, upper bounds. These bands encapsulate the effects of the geometry and ionization models with (orange) and without (blue) the inclusion of outflows for a MMD. As shown in the figure, even before accounting for the outflow phenomenology, the width of the uncertainty band spans over two orders of magnitude, approximately between 1 and 100 $M_\odot$, for~$f_\mathrm{PBH}\simeq 1$. This means that the intrinsic theoretical error in the modelling of the accretion makes these CMB anisotropy constraints unable to discern whether PBHs might be the progenitor of any observed BH merger, as they could either completely exclude or allow the LVK mass range. Once outflows are included in the modelling, the uncertainty band significantly enlarges both at large PBH masses, where MF dominates and the effect of non-thermal emission is negligible, and at low PBH masses, where the effect of non-thermal emission can be sizeable. Furthermore, the uncertainty due to the theoretical modelling in the case of EMDs can be easily seen when comparing the different panels of figure~\ref{fig:emd_bounds}.

In this respect, we also note that the existing bounds on the PBH abundance deriving from the LVK estimate of the binary BH local merger rate are subject of an ongoing debate. In fact, semi-analytical estimates of the local merger rate~$R_0$ of PBH binaries formed at early times suggest that it could be as high as~$R_0\sim 10^5\ \mathrm{Gpc^{-3}yr^{-1}}$~\cite{sasaki:pbhasdarkmatter, alihaimoud:pbhmergerrate} for~$f_\mathrm{PBH}=1$, to be compared with the observational value of~$R_0\sim 20\ \mathrm{Gpc^{-3}yr^{-1}}$~\cite{abbott:O3eventsproperties}. On the other hand, recent (and more accurate) numerical simulations reconcile the PBH early binary local merger rate value with the observed one for $f_{\rm PBH}=1$~\cite{jedamzik:earlybinariesmergerrateI, jedamzik:earlybinariesmergerrateII}.

The question of whether PBHs can be a significant component of the DM content of the universe and of the GW events detected by LVK is therefore still open and only a synergistic effort between different communities will help us find an answer to it. In fact, although in this paper we focused only on CMB constraints, they are not the only relevant ones in the LVK mass range. For instance, there are many complementary constraints coming from supernova lensing~\cite{zumalacarregui:pbhconstraints}, dwarf and ultra-faint dwarf galaxy dynamics~\cite{brandt:pbhconstraints, koushiappas:pbhconstraints, zoutendijk:pbhconstraints}, Lyman-$\alpha$ forest~\cite{afshordi:pbhconstraints,Murgia:2019duy} or wide binaries survival~\cite{monroyrodriguez:pbhconstraints}, which might help us shed light on the complexities of PBH phenomenology and ultimately determine their abundance.


\section{Conclusions}
\label{sec: concl}

Despite its remarkable success in explaining numerous cosmological observables, the~$\Lambda$CDM model cannot provide any insight on what the true nature of DM is, allowing a vast plethora of models to fit observations. For the sake of convenience, DM candidates are typically categorized in terms of their mass: already in this context, DM models span many orders of magnitude, ranging from ultralight axion-like particles to macroscopic compact objects like BHs. Hence finding novel ways to constrain DM properties becomes fundamental in order to establish its nature. 

In this work we focus on one of these popular candidates, PBHs. In particular, we focus on PBHs with masses larger than~$1\ M_\odot$, since they have the potential to be detected by existing and future GW observatories. This class of PBHs had been previously thought to be ruled out by existing LVK constraints on the local merger rate, although recent numerical simulations have shown that the number of PBH merging binaries had been overestimated by several orders of magnitude, reopening the possibility for PBHs to be a substantial component of the DM in the LVK mass range.

In this spirit, we turn our attention to other existing constraints on the PBH abundance in that same mass range, that is to those coming from the accretion of matter into a PBH. In fact, the emission of radiation following the accretion process affects the thermal history of the universe, by delaying recombination and anticipating reionization, and can therefore be constrained by CMB observations. However, large theoretical uncertainties underlay the modelling of accretion and this translates in large error bars on the final constraints. Examples of sources of such uncertainties are the geometry of the accretion as well as the ionization model determining the temperature profile close to the~BH.

On top of these uncertainties, one aspect of the accretion physics that has not been considered extensively in the literature so far is the effect of outflows (winds and/or jets, depending on their degree of collimation) on the accretion. However, it has been shown both analytically and numerically that even relatively weak outflows can escape the BH sphere of influence and sweep away at least part of the cosmic medium around the PBH, thereby decreasing its accretion rate. At the same time, the very same outflows could also accelerate non-thermal particles and effectively enhance the luminosity of the BH. Therefore, these competing effects can significantly affect the total luminosity of the system and introduce an additional layer of uncertainty that needs to be taken into account when quoting cosmological constraints derived from PBH accretion.

In this work we attempt to model the largely unknown nature of these effects and to analyse how their balance affects the CMB constraints on PBH accretion. We do so for different choices of accretion geometry (spherical or disk) and ionization models (photo- and collisional ionization), as well as for both monochromatic and extend mass distributions. Our quantitative findings are shown in figures~\ref{fig: bounds}-\ref{fig:emd_bounds}, which clarify that the final outcome heavily relies on the choice made in particular for the efficiency of the non-thermal emissions. This conclusion is graphically summarized in figure~\ref{fig: bounds_v2}, where the cumulative uncertainty bands on the ``true'' bounds are shown with and without the inclusion of outflows.

Looking towards the future, several developments might have a significant impact on our results. On the one hand, should strongly compelling arguments be put forward in favour of one particular accretion geometry, it would significantly reduce the uncertainty on the PBH abundance upper bounds reported in figure~\ref{fig: bounds_v2}. The same would be true also in the context of the ionization model. On the other hand, however, even more accurate simulations than the ones performed in \cite{Bosch-Ramon:2020pcz, Bosch-Ramon:2022eiy} might find indications for more complex outflow dynamics requiring, for instance, also information on outflow orientation or the transition to trans- or sub-sonic regimes. Taking into account for the potential role of these unknowns would inevitably further widen the size of the aforementioned uncertainty band. Similarly, the disk accretion scenario considered here is only restricted to the collisional ionization case, and extending it to the photo-ionization model would extend the uncertainty region towards low PBH masses.

Overall, we conclude that the path towards a realistic estimate of whether PBHs can make up for a sizeable fraction of the DM and of the events observed at the LVK facilities at the same time (i.e., in the same mass range) is still very long, as both the cosmological constraints and the merger rate estimates are subject to very large theoretical uncertainties. While this should discourage premature claims of exclusion or detection, it should also be seen as a source of motivation for the improvements to come.


\section*{Acknowledgements}
The authors sincerely thank Yacine Ali-Ha\"imoud, Sebastien Clesse and Pasquale Serpico for the useful comments and Federico Mussetti for the support. ML is supported by an F.R.S.-FNRS fellowship and by the IISN convention 4.4503.15. NB acknowledges partial support from the National Science Foundation (NSF) under Grant No.~PHY-2112884. VB-R acknowledges financial support from the State Agency for Research of the Spanish Ministry of Science and Innovation under grant PID2019-105510GB-C31. V.B-R. is Correspondent Researcher of CONICET, Argentina, at the IAR. AR acknowledges funding from Italian Ministry of University and Research (MUR) through the ``Dipartimenti di eccellenza'' project ``Science of the Universe''. 
LV acknowledges support by project PGC2018-098866-B-I00 MCIN/AEI/10.13039/501100011033 y FEDER “Una manera de hacer Europa” and European Union’s Horizon 2020 research and innovation programme ERC (BePreSysE, grant agreement 725327). This work was supported by the State Agency for Research of the Spanish Ministry of Science and Innovation through the ''Unit of Excellence Mar\'ia de Maeztu 2020-2023'' award to the Institute of Cosmos Sciences (CEX2019-000918-M funded by MCIN/AEI /10.13039/501100011033). Computational resources have been provided by the Consortium des Équipements de Calcul Intensif (CÉCI), funded by the Fonds de la Recherche Scientifique de Belgique (F.R.S.-FNRS) under Grant No. 2.5020.11 and by the Walloon Region.

\appendix
\section{Converting abundance constraints from MMD to EMD}
\label{app:emd_converting_relations}

Abundance constraint conversion formulas accounting for PBH effects on CMB anisotropies exist only for the spherical accretion case~\cite{Bellomo:2017zsr,2017JCAP...10..052B}. However, different accretion geometries or energy emission efficiencies require the development (and testing) of new conversion relations, which we present in this appendix. Furthermore, in this appendix we focus on lognormal EMDs, given the large interest on this specific EMD coming from the theoretical modelling of PBH formation. We note, however, that the procedure presented in this appendix is easily applicable also to other EMDs. 

In the most general setup, as for instance in equation~\eqref{eq: L_tot}, we have that the injected and deposited energy are proportional to
\begin{equation}
    \frac{L_\mathrm{tot}}{M_\mathrm{PBH}} = \left(\epsilon_\mathrm{non-th} + \epsilon_\mathrm{rad}\right) \frac{\dot{M}_\mathrm{PBH}}{M_\mathrm{PBH}},
\end{equation}
which is weighted by the PBH EMD of choice, as shown in equation~\eqref{eq:injected_energy_emd}. In the spherical accretion case we have that
\begin{equation}
    \frac{L_\mathrm{tot}}{M_\mathrm{PBH}} \propto \left[\epsilon_\mathrm{non-th} + \gamma_\mathrm{sph} \lambda M_\mathrm{PBH}\right] \lambda M_\mathrm{PBH},
\end{equation}
where in the redshift range of interest the mass-independent quantity~$\gamma_\mathrm{sph}$ takes values in the range~$\gamma_\mathrm{sph} \simeq \left[2-19\right]\times 10^{-10}$ and~$\gamma_\mathrm{sph} \simeq \left[3-24\right]\times 10^{-8}$ for the collisional and photoionization models, respectively. The effect of the dimensionless accretion rate can be effectively parametrized (neglecting its redshift dependence) as~$\lambda\propto M^{\alpha/2}$~\cite{Bellomo:2017zsr}, where $\alpha$ is a parameter to be tuned numerically a posteriori to minimize the differences in the relevant observable quantity between the EMD case and the equivalent monochromatic case. Therefore, for the purpose of obtaining an accurate conversion, the equivalent mass is given by
\begin{equation}
    \left[\epsilon_\mathrm{non-th} + \gamma_\mathrm{sph} M_\mathrm{eq}^{1+\alpha/2}\right] M^{1+\alpha/2}_\mathrm{eq} = \mu^{1+\alpha/2} e^{(2+\alpha)^2\sigma^2/8} \left[\epsilon_\mathrm{non-th} + \gamma_\mathrm{sph} \mu^{1+\alpha/2} e^{3(2+\alpha)^2\sigma^2/8} \right],
\end{equation}
where, for a lognormal distribution, $\alpha = 0.2$. From the equation above we see that in the limit of zero non-thermal emission~$(\epsilon_\mathrm{non-th} \to 0)$ we recover the known result~\cite{Bellomo:2017zsr}
\begin{equation}
    M_\mathrm{eq}^{2+\alpha} = \mu^{2+\alpha} e^{(2+\alpha)^2\sigma^2/2},
\label{eq:M_eq_spharical}
\end{equation}
while in the limit of dominant non-thermal emission~$(\gamma_\mathrm{sph} \to 0)$, we find that
\begin{equation}
    M^{1+\alpha/2}_\mathrm{eq} = \mu^{1+\alpha/2} e^{(2+\alpha)^2\sigma^2/8}.
\end{equation}

On the other hand, in the case of disk accretion we have
\begin{equation}
    \frac{L_\mathrm{tot}}{M_\mathrm{PBH}} \propto \left[\epsilon_\mathrm{non-th} + \gamma_\mathrm{disk} M_\mathrm{PBH}^a\right] M_\mathrm{PBH},
\end{equation}
where in this case the mass-independent quantity~$\gamma_\mathrm{disk} \simeq 6 \times 10^{-4}$ and the exponent~$a$ is given in ref.~\cite{Xie:2012rs}. In this case the equivalent mass relation reads as
\begin{equation}
    \left[\epsilon_\mathrm{non-th} + \gamma_\mathrm{disk} M_\mathrm{eq}^{a} \right] M_\mathrm{eq} = \mu \left[ \epsilon_\mathrm{non-th} e^{\sigma^2/2} + \gamma_\mathrm{disk} \mu^{a} e^{(1+a)^2 \sigma^2/2} \right]
\label{eq:M_eq_disk}
\end{equation}
which reduces to 
\begin{equation}
    M_\mathrm{eq}^{1+a} = \mu^{1+a} e^{(1+a)^2\sigma^2/2},
\end{equation}
and 
\begin{equation}
    M_\mathrm{eq} = \mu e^{\frac{\sigma^2}{2}}
\end{equation}
in the radiation-emission and non-thermal-emission dominated regimes, respectively.

\bibliography{bibliography}
\bibliographystyle{utcaps}

\end{document}